\documentstyle[12pt]{article}

\input epsf.sty

\newcommand{\res}{\mathop{\rm Res}}
\newcommand{\g}{{\sl g}}

\newcommand{\X}{{\cal X}}
\newcommand{\D}{{\cal D}}
\newcommand{\K}{{\cal K}}
\newcommand{\G}{{\cal G}}
\newcommand{\W}{{\cal W}}
\newcommand{\V}{{\cal V}}
\newcommand{\I}{{\cal I}}

\begin{document}
\begin{titlepage}

\centerline{\large \bf Next-to-leading order evolution of twist-two}
\centerline{\large \bf conformal operators: The Abelian case.}

\vspace{10mm}

\centerline{\bf A.V. Belitsky$^{a,b,}$\footnote{Alexander von
            Humboldt Fellow.}, D. M\"uller$^{a,c}$}

\vspace{18mm}

\centerline{\it ${^a}$Institut f\"ur Theoretische Physik, Universit\"at
                Regensburg}
\centerline{\it D-93040 Regensburg, Germany}
\centerline{\it ${^b}$Bogoliubov Laboratory of Theoretical Physics,
                Joint Institute for Nuclear Research}
\centerline{\it 141980, Dubna, Russia}
\centerline{\it ${^c}$Institute of Theoretical Physics, Leipzig
                University}
\centerline{\it  04109 Leipzig, Germany}

\vspace{20mm}

\centerline{\bf Abstract}

\hspace{0.8cm}

\noindent
We present the method, based on the use of the broken conformal Ward
identities, for the calculation of the anomalous dimensions of conformal
operators beyond the leading order of perturbation theory. By means of
this technique we find the complete set of two-loop corrections to
the off-diagonal elements of the anomalous dimension matrix (the
diagonal ones are known from deep inelastic scattering calculations)
of the parity odd and even singlet local operators in the Abelian gauge
theory. We reconstruct corresponding exclusive two-loop evolution
kernels and give the reduction formulae for the restoration of the
diagonal part of the ER-BL kernels from the DGLAP splitting functions.

\vspace{0.5cm}


\vspace{0.5cm}


\end{titlepage}

\section{Introduction.}

Conformal invariance \cite{Kastrup62,MackSalam69,FGG73} has been
known for a long time as a powerful tool for the derivation of the
constraints on the form of the Green functions in massless field
theory. However, the real world could not be conformal invariant
even approximately, since otherwise the mass spectrum of the
elementary particles might not be discrete. When the interaction
is switched on the necessity to construct well-defined theory
forces us to introduce a UV cut-off, which inevitably breaks the
symmetry. It is reflected in the fact that the field operators
acquire the anomalous dimensions and the coupling constant becomes
running. In the description of the physical quantities new objects
--- local composite operators --- appear which define the dynamical
content of the former. The computation of the spectrum of the
anomalous dimensions of these composite operators is one of the most
important problems of the theory, since these quantities determine
the scaling behaviour of experimental observables, i.e. the moments
of the parton distributions or wave functions depending on the
particular processes in question. These non-perturbative entries
satisfy corresponding evolution equations, known as
Dokshitzer-Gribov-Lipatov-Altarelli-Parisi (DGLAP)
\cite{Lip72,Alt77,Dok77} one for inclusive reactions and
Efremov-Radyushkin-Brodsky-Lepage (ER-BL) equation \cite{ER,BL} for
the exclusive ones \cite{BL89}. In order to solve them one has to
determine the eigenfunctions and eigenvalues of renormalization
group kernels. The main problem arises in the study of exclusive
processes \cite{ER} where contrary to the forward scattering one
needs to handle the whole tower of local operators with total
derivatives which mix with each other under renormalization.
In the leading order (LO) this problem was analyzed in a number
of papers \cite{ER,BL,BFLS80,Cha80,Ohr80,Mak81,Ohr82,BDFL86}.
It was recognized that the operators which diagonalize the anomalous
dimension matrix to the lowest order have a conformally covariant
transformation behaviour at tree level.

Nowadays it becomes a common wisdom to perform the analyses in the
next-to-leading order (NLO) of perturbation theory. The self-consistent
approach to this problem requires the knowledge of the two-loop exclusive
evolution kernels. Because of the complications spelled out above, the
direct calculation of the anomalous dimension matrices or the evolution
kernels beyond the leading order turns out to
be tremendous task and was resolved up to now in the non-singlet sector
\cite{DitRad84,Sam84,MikhRad85,Katz85,MikhRad86,Mue88,MRGDH94}. Since the
conformal operators diagonalize corresponding ER-BL evolution kernels
in the leading order of perturbation theory and the off-diagonal part
appears only in the NLO approximation, the symmetry breaking
starts at the order ${\cal O}(\alpha)$ and might be expected to be
evaluated from the Green functions of some quasi-one-loop objects.
Therefore, the first step toward the accomplishment of the program of
the evaluation of the ${\cal O}(\alpha^2)$-corrections to the evolution
kernels, is to specify the particular pattern of the conformal symmetry
breaking in gauge theories to leading order. To this end true conformal
Ward identities are required. Due to the UV divergences of the theory
there exist anomalies, but only part of them could be related to those
which appear in the violation of scale invariance \cite{Jackiw70}. Note,
that this is already true in scalar theories, where there is one-to-one
correspondence between dilatation and special conformal anomalies
induced by the renormalization of the fields \cite{Symanzik71} only.
In gauge theories, a priori, one expects in addition a breaking due to
the gauge fixing and ghost terms in the Lagrangian, but it should not
affect the final result provided we are interested in some physically
relevant quantities. Thus, the limit of non-running coupling constant
corresponds to the exact scale invariance; however, it is by no means
true that the special conformal symmetry also respects this limit
because there exists a new source of symmetry breaking which can be
attributed to the renormalization of the product of composite operators
and trace anomaly of the improved energy-momentum tensor. However, in
the study of gauge invariant quantities these effects could be always
embedded into the redefined conformal representation of the algebra, so
that the symmetry will be broken explicitly only by the terms
proportional to the $\beta$-function of the theory. Thus, conformal
covariance is restored in the non-trivial fixed point of the
renormalization group function. This fact gives us the opportunity
to apply the powerful machinery of conformal symmetry. It has been
exploited by us quite recently to give several QCD predictions
coming from the use of the conformally covariant operator product
expansion \cite{Mul97,BM97a,BS98-1}.

In the present analysis we will construct the formalism and calculate
the two-loop results for the singlet ER-BL kernels, restricting ourselves
to Abelian gauge theory. The main tools for these purposes will be the
renormalized Ward identities for the dilatation and special
conformal transformations \cite{Nielsen73,Sarkar74,Nielsen75,MulZ91,Mul94}
and the commutator constraint coming from the algebra of the conformal
group. The most important feature of the former is that they are derived
without any assumptions about the possible form of the symmetry breaking
and, thus, turn out to be true anomalous Ward identities in gauge theory.

Consequent presentation will be organized as follows. Next section is
mainly introductory. We summarize there some basic definitions of the
main objects of our analysis: singlet conformal operators, renormalized
Lagrangian of the Abelian theory, as well as other generalities related
to the transformation of the latter under the action of generators of
the conformal algebra. In section 3 we construct renormalized conformal
Ward identities for the Green functions with conformal operator insertions.
In section 4 they are used to derive the commutator constraint on the
off-diagonal elements of the anomalous dimension matrix. The most
important feature of the conformal consistency relation is that it
allows to find corresponding entries in the $n$-th order from the
calculation of the, so-called, special conformal anomaly matrix in the
$(n-1)$-loop approximation. This is used by us to derive the two-loop
results for the anomalous dimensions in terms of the one-loop quantities
which are evaluated in section 5. These results are also transformed in
the language of ER-BL evolution kernels and checked in section 6
for the particular set of two-loop diagrams with fermion bubble
insertions in the $GQ$-kernels. Next, we present the reduction
formulae which allows to reconstruct the diagonal part of the ER-BL
kernels starting from DGLAP analogues. The final section is devoted to
the summary. Trying to make the paper self-consistent as much as
possible we add two appendices. In the first one we give the Feynman
rules used in the main body of the paper. In the second one we list
the extended ER-BL evolution kernels responsible for the evolution of
the generalized non-forward distribution functions which attract lots
of attention recently.

\section{Preliminaries.}

To start with, we review few basic properties of the theory invariant
under conformal algebra of space-time. The conformal algebra
\cite{Kastrup62,MackSalam69,FGG73} is a 15-dimensional Lie algebra,
isomorphic to $O(4,2)$. Its generators are the Poincar\'e generators
${\cal P}_\mu$, ${\cal M}_{\mu\nu}$ together with dilatation $\D$ and
special conformal transformation $\K_\mu$. The latter could be
defined as product $\K_\mu = {\cal R} {\cal P}_\mu
{\cal R}^{-1}$, where ${\cal R}$ produces inversion of space-time
points $x_\mu \to x_\mu/ x^2 $. Apart from commutators which
manifest the scalar and vector nature of $\D$ and $\K_\mu$,
respectively, there are extra non-trivial commutation relations
\begin{equation}
\left[ \D, \K_\mu \right]_- = i \K_\mu ,
\quad
\left[ {\cal P}_\mu, \D \right]_- = i {\cal P}_\mu ,
\quad
\left[ \K_\mu, {\cal P}_\nu \right]_-
= - 2 i \left[
g_{\mu\nu} \D + {\cal M}_{\mu\nu}
\right].
\end{equation}
The transformations of the field operators under dilatation and special
conformal trans\-for\-ma\-tions\footnote{The $\delta$-sign means the
variation of the shape of the fields only $\delta \phi = \phi' (x) -
\phi (x)$.} are expressed by the formulae
\begin{equation}
\label{transformation}
\begin{array}{ll}
\delta^S x_\mu = x_\mu, &\delta^S \phi (x)
= i \left[ \phi (x), \D \right]_- = i \hat D \phi (x), \\
\delta_\nu^C x_\mu = 2x_\mu x_\nu - x^2 g_{\mu\nu},
&\delta_\mu^C \phi (x)
= i \left[ \phi (x), \K_\mu \right]_- = i \hat K_\mu \phi (x) ,
\end{array}
\end{equation}
where the  parameters of the infinitesimal transformations have been
set equal to unity. Using the method of induced representation
\cite{MackSalam69} one can construct from the, so-called, stability
subgroup the differential representation $\hat G$ for the generators
$\D$ and $\K_\mu$ of the conformal algebra. Here we deal with the
irreducible standard representation, namely:
\begin{equation}
\hat D = i \left( d_\phi + x_\mu \partial_\mu \right) , \qquad
\hat K_\mu = i \left( 2d_\phi x_\mu - x^2 \partial_\mu
+ 2x_\mu x_\nu \partial_\nu - 2i x_\nu \Sigma_{\mu\nu} \right) ,
\end{equation}
where $d_\phi$ is a scale dimension of the field $\phi$. The particular
choice for them will be discussed below. The action of the spin operator
$\Sigma_{\mu\nu}$ on the fermion and vector fields is defined as usual:
\begin{equation}
\Sigma_{\mu\nu} \psi
= \frac{i}{4}\left[ \gamma_\mu, \gamma_\nu \right]_- \psi,
\quad
\Sigma_{\mu\nu} A_\rho
= i ( g_{\mu\rho} A_\nu - g_{\nu\rho} A_\mu ) .
\end{equation}

In this paper we intend to construct and study the Ward identities
which originates from the dilatation and special conformal
transformations. The simplest way to derive them is to use the
path integral formalism \cite{Sarkar74,Nielsen75,MulZ91,Mul94} and
dimensional regularization in $d = 4 - 2\epsilon$ dimensions. Their
generic form looks like
\begin{equation}
\label{WI}
\langle \delta^G [{\cal O}_{jl}] \X \rangle
+ \langle [{\cal O}_{jl}] \delta^G [iS] \X \rangle
+ \langle [{\cal O}_{jl}] \delta^G \X \rangle = 0,
\end{equation}
where $\delta^G$-symbol means some infinitesimal transformation from the
algebra and $\langle \dots \rangle $ corresponds to the path integral
averaging with the weight function $\exp i S$. $S=\int d^d x {\cal L}$
stands for the action defined in terms of the renormalized Lagrangian
of Abelian gauge fields coupled to fermions
\begin{equation}
{\cal L} =
- \frac{1}{4} Z_3 \left( G_{\mu\nu} \right)^2
- \frac{1}{2\xi} \left( \partial_\mu A_\mu \right)^2
+ Z_2 \bar \psi i \!\not\!\partial \psi
+ Z_1 \mu^\epsilon \g \bar \psi \!\not\!\!A \psi
+ \partial_\mu \bar\omega \partial_\mu \omega ,
\end{equation}
where $G_{\mu\nu} = \partial_\mu A_\nu - \partial_\nu A_\mu$ is the
Abelian field strength tensor. Here we have also added a free ghost
Lagrangian to make ${\cal L}$ invariant under BRST transformation.
In Eq. (\ref{WI}) $\X = \prod_k \phi (x_k)$ is a product of elementary
fields (entering in the Lagrangian); its variation is expressed in
terms of the generators $\G$ of the conformal group acting on the
field operators:
\begin{equation}
\delta^G \X = \sum_{i = 1}^{k} \phi (x_1)\phi (x_2) \dots
i \left[ \phi (x_i), \G \right]_- \dots \phi (x_k).
\end{equation}
In Eq. (\ref{WI}) we have considered Green functions with a conformal
operator insertion which is defined as two-dimensional vector
\begin{equation}
{\cal O}_{jl}
= \left( { {^Q\!{\cal O}_{jl}} \atop {^G\!{\cal O}_{jl}} } \right) .
\end{equation}
An explicit form of the conformal covariant tree operators
is\footnote{Throughout the paper we use the conventions adopted by
Itzykson and Zuber \cite{itz80} for Dirac $\gamma$-matrices and
Lorentz tensors.}
\begin{equation}
\label{treeCO}
\left\{\!\!\!
\begin{array}{c}
{^Q\!{\cal O}^V} \\
{^Q\!{\cal O}^A}
\end{array}
\!\!\!\right\}_{jl}
\!=
\bar{\psi} (i \partial_+)^l\!
\left\{\!\!\!
\begin{array}{c}
\gamma_+ \\
\gamma_+ \gamma_5
\end{array}
\!\!\!\right\}
\!C^{\frac{3}{2}}_j\!
\left( \frac{\stackrel{\leftrightarrow}{D}_+}{\partial_+} \right)
\!\psi, \
\left\{\!\!\!
\begin{array}{c}
{^G\!{\cal O}^V} \\
{^G\!{\cal O}^A}
\end{array}
\!\!\!\right\}_{jl}
\!=
G_{+ \mu} (i \partial_+)^{l-1}\!
\left\{\!\!\!
\begin{array}{c}
g_{\mu\nu} \\
i\epsilon_{\mu\nu-+}
\end{array}
\!\!\!\right\}
\!C^{\frac{5}{2}}_{j - 1}\!
\left(
\frac{\stackrel{\leftrightarrow}{\partial}_+}{\partial_+}
\right)
\!G_{\nu +},
\end{equation}
where $\partial \!= \stackrel{\rightarrow}{\partial}
\!\!+\!\! \stackrel{\leftarrow}{\partial}$ and
$\stackrel{\leftrightarrow}{D} = \stackrel{\rightarrow}{\partial}\!\!
-\!\! \stackrel{\leftarrow}{\partial} \!\!- 2i \mu^\epsilon \g A$.
We have used the following relation $\epsilon_{\mu\nu-+} G_{+\mu}^{(2)}
G_{\nu +}^{(1)} = G_{+\mu}^{(2)} \widetilde G_{\mu +}^{(1)} = -
\widetilde G_{+\mu}^{(2)} G_{\mu +}^{(1)}$. The renormalized operators
are expressed through the non-renormalized ones in the following way
\begin{equation}
[ {\cal O}_{jl} ]
=
\sum_{k = 0}^{j} Z_{jk}(\epsilon, \g) {\cal O}_{kl} ,
\end{equation}
where the matrix $Z_{jk}(\epsilon, \g)$ is triangular and independent
on $l$ due to Lorentz invariance of the theory. It is given in the
minimal subtraction scheme by
\begin{eqnarray}
\hat Z =
\left({
{^{QQ}\!\hat Z}\ {^{QG}\!\hat Z}\atop
{^{GQ}\!\hat Z}\ {^{GG}\!\hat Z}
}\right) \quad \mbox{with\ \ }
\hat Z = \hat 1 + \frac{1}{\epsilon} \hat Z^{[1]} +\cdots.
\end{eqnarray}

Since we are restricted to the leading twist operators (\ref{treeCO})
defined on the light-cone, it is sufficient to deal consequently with
the collinear conformal subgroup $SU( 1, 1 )$ of the whole group which
simplify considerably the analysis. For that we project the Lorentz
indices with two light-like vectors $n$ and $n^*$, which are defined
along opposite tangents to the light cone, with $n^2 =
n^{*2} = 0$ and $nn^* = 1$.

Beyond the tree level we are not limited to any unique choice of the
scale dimensions $d_\phi$, rather we can define them as we wish. It
turns out, that the most optimal and convenient set is $d_G = 1$,
$d_\psi = {\scriptstyle\frac{3}{2}}$. Note, that the Ward identities
are invariant with respect to change of the scale dimensions, since
the generating functional is invariant under linear transformation
of the field operators. For the scale dimensions of the fields we
have chosen, the variations of the conformal operators are very
simple\footnote{We have used the covariant conformal transformation
for the gauge field strength tensor given in Ref. \cite{Jackiw78}.}:
\begin{equation}
\delta^S \, [{\cal O}_{jl}]
= - ( l + 3 ) [{\cal O}_{jl}] , \qquad
\delta^C_- \, [{\cal O}_{jl}]
= i\, \sum_{k = 0}^{j} \left\{ \hat Z \,
\hat a (l) \hat Z^{-1} \right\}_{jk}
[{\cal O}_{k l-1}] ,
\end{equation}
where $a_{jk}(l) = \delta_{jk} a (j,l) = \delta_{jk} 2
\{ ( j + 1 )( j + 2 ) - ( l + 1 )( l + 2 ) \} $.

In the consequent discussion and all practical calculations we will
always employ the light-cone position formalism for the evaluation
of the renormalization matrices of different composite operators.
Therefore, they should be replaced for these purposes by the non-local
string operators. Thus, we introduce the quark and gluon gauge invariant
operators
\begin{eqnarray}
\label{string-operator}
\left\{\!\!\!
\begin{array}{c}
{^Q\!{\cal O}^V} \\
{^Q\!{\cal O}^A}
\end{array}
\!\!\!\right\}
(\kappa_1, \kappa_2)
&=&
\bar{\psi} (\kappa_2 n)
\left\{\!\!\!
\begin{array}{c}
\gamma_+ \\
\gamma_+ \gamma_5
\end{array}
\!\!\!\right\}
\!\psi (\kappa_1 n)
\mp
\bar{\psi} (\kappa_1 n)
\left\{\!\!\!
\begin{array}{c}
\gamma_+ \\
\gamma_+ \gamma_5
\end{array}
\!\!\!\right\}
\!\psi (\kappa_2 n) , \\
\left\{\!\!\!
\begin{array}{c}
{^G\!{\cal O}^V} \\
{^G\!{\cal O}^A}
\end{array}
\!\!\!\right\}
(\kappa_1, \kappa_2)
&=&
G_{+ \mu} (\kappa_2 n)
\left\{\!\!\!
\begin{array}{c}
g_{\mu\nu} \\
i\epsilon_{\mu\nu-+}
\end{array}
\!\!\!\right\}
\!G_{\nu +} (\kappa_1 n),
\end{eqnarray}
where we have omitted for brevity the path ordered exponential in
the first entry.

To simplify the derivation of the Ward identities it is instructive
to use the following variations of the Lagrangian
\begin{eqnarray}
\label{D-variation}
\delta^S {\cal L} &=&
- \partial_\mu \left( x_\mu {\cal L} \right)
+ d \, {\cal L}
- d_\phi \, \frac{\partial{\cal L}}{\partial\phi}\phi
- (d_\phi + 1) \, \frac{\partial{\cal L}}{\partial(\partial_\mu\phi)}
\partial_\mu \phi ,\\
\delta_\nu^C {\cal L} &=&
- \partial_\mu
\left( \left( 2x_\mu x_\nu - x^2 g_{\mu\nu} \right) {\cal L} \right)
+ 2 x_\nu\left(
d \, {\cal L}
- d_\phi \, \frac{\partial{\cal L}}{\partial\phi}\phi
- (d_\phi + 1) \, \frac{\partial{\cal L}}{\partial(\partial_\mu\phi)}
\partial_\mu \phi \right) \nonumber\\
&-& 2 d_\phi \, \frac{\partial{\cal L}}{\partial(\partial_\nu\phi)}
\phi
+ 2i \frac{\partial{\cal L}}{\partial(\partial_\mu\phi)}
\Sigma_{\nu\mu} \phi .
\label{C-variation}
\end{eqnarray}
They lead to the non-renormalized scale and special conformal variations
of the action
\begin{eqnarray}
\label{var-S}
\delta^S S &=& \epsilon \int d^d x
\left\{
{\cal O}_A (x) + {\cal O}_B (x) - {\mit \Omega}_{\bar\psi\psi} (x)
\right\}, \\
\label{var-C}
\delta_-^C S &=& \int d^d x 2 x_-
\left\{
\epsilon
\left[
{\cal O}_A (x) + {\cal O}_B (x) - {\mit \Omega}_{\bar\psi\psi} (x)
\right]
+ (d - 2) \partial_\mu {\cal O}^\mu_B (x)
\right\} ,
\end{eqnarray}
expressed in terms of the following equation-of-motion
(EOM) operators
\begin{equation}
{\mit \Omega}_G (x) = A_\mu \frac{\delta S}{\delta A_\mu},\quad
{\mit \Omega}_{\bar\psi} (x)
= \bar\psi \frac{\delta S}{\delta \bar\psi}
\quad \mbox{and}
\quad
{\mit \Omega}_\psi (x) = \frac{\delta S}{\delta \psi} \psi ,
\end{equation}
(${\mit \Omega}_{\bar\psi\psi} = {\mit \Omega}_{\psi}
+ {\mit \Omega}_{\bar\psi}$), and the operators of class $A$ and $B$
\cite{LeeJoglekar75}
\begin{eqnarray}
{\cal O}_A (x) = \frac{Z_3}{2} \left( G_{\mu\nu} \right)^2, \ \
{\cal O}_B (x) = \frac{\delta^{\rm BRST}}{\delta\lambda}
\bar\omega\partial_\mu A_\mu, \ \
{\cal O}^\nu_B (x) = \frac{\delta^{\rm BRST}}{\delta\lambda}
\bar\omega A^\nu .
\end{eqnarray}
Because of the factor $\epsilon$ appearing in Eqs.
(\ref{var-S},\ref{var-C}), the conformal variation of the action can be
different from zero only due to the UV divergencies. The appearing
operators can be renormalized along the line of Refs. \cite{LeeJoglekar75}.
However, since these are exactly the operators which enter in the
Lagrangian density, their renormalization could be gained from the
study of the differential vertex operator insertion
${\scriptstyle\g}\frac{\partial}{\partial\g} S$ and ${\scriptstyle\xi}
\frac{\partial}{\partial\xi} S$. Namely, a little algebra leads to
\begin{equation}
[\Delta^\g]
= \left[{\cal O}_A\right]
+ \left[{\cal O}_B\right]
+ {\mit \Omega}_G ,
\quad
[\Delta^\xi]
= \frac{1}{2} \left[ {\cal O}_B \right] .
\end{equation}
Here\footnote{For simplifying notations, we use the following
conventions: for unintegrated operator insertions and EOM operators
arisen form the conformal variations of the Lagrangian we keep the
dependence on the space-time point explicit $\Delta (x) $; for the
integrated quantities with weight function 1 we just omit this
dependence $\Delta \equiv \int d^d x \Delta (x)$, while for weight
$2x_-$ we use $\Delta^- \equiv \int d^d x \, 2 x_- \Delta (x)$.}
\begin{equation}
[\Delta^\g (x)] \equiv \g \frac{\partial}{\partial\g} {\cal L} ,
\quad
[\Delta^\xi (x)] \equiv \xi \frac{\partial}{\partial\xi} {\cal L}
\quad \mbox{and}
\quad
[\Delta^{\rm ext} (x)] \equiv \partial_\mu [{\cal O}_B^\mu (x) ] .
\end{equation}
From the finiteness of the differential vertex operator insertions
the counterterms of $[ {\cal O}_A (x) ]$ and $[ {\cal O}_B (x) ]$
can be fixed:
\begin{eqnarray}
\left[ {\cal O}_A (x) \right]
&=&
\left( 1 - \frac{1}{2} \g \frac{\partial\ln Z_3}{\partial \g} \right)
{\cal O}_A (x)
+ \frac{1}{2}
\left( \g \frac{\partial\ln Z_2}{\partial \g}
- 2\xi\frac{\partial\ln Z_2 }{\partial \xi}\right)
{\mit \Omega}_{\bar\psi\psi} (x) , \\
\left[ {\cal O}_B (x) \right]
&=&
{\cal O}_B (x)
+ \xi\frac{\partial\ln Z_2 }{\partial \xi}\
{\mit \Omega}_{\bar\psi\psi} (x)
,\nonumber
\end{eqnarray}
in complete agreement with general renormalization theorems
\cite{LeeJoglekar75}. These equalities hold up to counterterms which
are total derivatives of some other operators ${\cal R}_i^\mu$. By
considering the Green functions of the differential operator insertion
${\scriptstyle\xi} \frac{\partial}{\partial\xi} S$ and from the
finiteness of the special conformal Ward identity, we can conclude that
they are proportional to the gauge fixing parameter $\xi$ and thus
vanish identically in Landau gauge. The operator ${\cal O}_B^\mu
= [{\cal O}_B^\mu]$ is finite by itself, since there is no other
dimension three operator with the same quantum numbers which it could
mix with due to the renormalization. Now the final step can be easily
done with the result
\begin{eqnarray}
\delta^S S &=&
- \frac{\beta_\epsilon}{\g}
[ \Delta^\g ]
- \sigma [ \Delta^\xi ]
+ ( \gamma_G - \epsilon ) {\mit \Omega}_G
+ ( \gamma_\psi - \epsilon ) {\mit \Omega}_{\bar\psi\psi} ,
\nonumber\\
\delta^C_- S &=&
- \frac{\beta_\epsilon}{\g}
[ \Delta^\g_- ]
- \sigma [ \Delta^\xi_- ]
+ (d - 2) [ \Delta^{\rm ext}_- ]
+ ( \gamma_G - \epsilon ) {\mit \Omega}_G^-
+ ( \gamma_\psi - \epsilon ) {\mit \Omega}_{\bar\psi\psi}^- ,
\end{eqnarray}
where the coefficients in front of the operator insertions are
the well-known renormalization group functions
\begin{eqnarray}
\beta_\epsilon (\g ) = \mu \frac{d \g }{d \mu}
= - \epsilon \, \g + \beta (\g ),
\quad
\sigma = \mu \frac{d}{d \mu} \ln \xi
\end{eqnarray}
($\sigma = -2 \gamma_G$) and the anomalous dimensions of the
elementary fields
\begin{eqnarray}
\gamma_\psi = \frac{1}{2}\mu\frac{d}{d\mu} \ln Z_2 ,
\quad
\gamma_G = \frac{1}{2}\mu\frac{d}{d\mu}\ln Z_3.
\end{eqnarray}
The first expansion coefficients of $\gamma_i =
\frac{\alpha}{2\pi} \gamma^{(0)}_i + \dots $ and $\frac{\beta
(\g)}{\g} = \frac{\alpha}{4 \pi} \beta_0 + \dots$ in perturbative
series are
\begin{equation}
\beta_0 = \frac{4}{3}T_F N_f,
\qquad \gamma^{(0)}_\psi = \frac{\xi}{2} C_F,
\qquad \gamma^{(0)}_G = \frac{2}{3}T_F N_f.
\end{equation}
Aiming consequent generalization of the present considerations to
the non-Abelian theory we keep dependence on the colour factors
explicit. In QED we have to set $C_F = T_F = 1$.

\section{Conformal Ward identities.}

Combining the results, we have obtained so far in Eq. (\ref{WI}), we
will come to conformal Ward identities, which, however, contain
unrenormalized products of operators. The aim of this section is to
renormalize these products \cite{Joglekar77-1}. Since the coefficients
appearing in the Ward identities depend on the regularization parameter
$\epsilon$, this renormalization procedure leads to the conformal
anomalies.

\subsection{Scale Ward identity.}

The renormalization of the dilatation Ward identity is quite
straightforward and does not require any additional information
as one which comes from the consideration of the differential
vertex insertions.
By using the action principle \cite{CollinsBOOK} we can get the
results for the products $[{\cal O}_{jl}][\Delta^i]$ from the Green
functions with the conformal operator insertion $[{\cal O}_{jl}]$:
\begin{eqnarray}
i [ {\cal O}_{jl} ] [\Delta^\g ]
=
i [ {\cal O}_{jl} \, \Delta^\g ]
\!\!&-&\!\!
\sum_{k = 0}^{j}
\biggl\{
\g \frac{\partial \hat Z}{\partial \g} \hat Z^{-1}
+
\hat Z
\left( \hat 1
\int d^d x A_\mu (x) \frac{\delta}{\delta A_\mu (x)}
-
2 \hat P_G
\right) \hat Z^{-1}
\biggr\}_{jk}
[ {\cal O}_{kl} ] , \\
i [ {\cal O}_{jl} ] [\Delta^\xi ]
=
i [ {\cal O}_{jl} \, \Delta^\xi ]
\!\!&-&\!\!
\sum_{k = 0}^{j}
\biggl\{
\xi \frac{\partial \hat Z}{\partial \xi} \hat Z^{-1}
\biggr\}_{jk}
[ {\cal O}_{kl} ] .
\end{eqnarray}
When substituted into the Ward identity, the terms proportional to the
functional derivative acting on the conformal operator will cancel
upon similar terms coming from the renormalization of $[{\cal O}_{jl}]
{\mit \Omega}_G$:
\begin{equation}
i [{\cal O}_{jl}] {\mit \Omega}_\phi
= i [{\cal O}_{jl} {\mit \Omega}_\phi ]
- \int d^d x \phi (x) \frac{\delta}{\delta \phi (x)} [{\cal O}_{jl}].
\end{equation}
For quark equation of motion we have
\begin{equation}
\int d^d x \left(
\bar\psi \frac{\delta}{\delta \bar\psi}
+
\frac{\delta}{\delta \psi} \psi
\right)
[{\cal O}_{jl}]
= 2 \sum_{k = 0}^{j}
\left\{
\hat Z \hat P_Q \hat Z^{-1}
\right\}_{jk} [{\cal O}_{kl}],
\end{equation}
where the projectors on the quark and gluon sectors are defined by
\begin{eqnarray}
\hat P_Q =
\left( { 1 \ 0 \atop 0 \ 0 } \right),
\qquad
\hat P_G =
\left( { 0 \ 0 \atop 0 \ 1 } \right).
\end{eqnarray}
It should be noted that the definition of the renormalization constant
we have used above does not correspond to the minimal subtraction of
UV divergencies, rather it comes from the mere integration by parts in
the functional integral for the Green functions.

Putting all these equations together, we obtain the renormalized
dilatation Ward identity
\begin{eqnarray}
\label{DWI}
&&\langle [ {\cal O}_{jl} ] \delta^S \X \rangle
= \sum_{k = 0}^{j}
\left\{
\hat\gamma_{Z}
+ ( l + 3 ) \hat 1
+ 2 ( \gamma_G - \epsilon ) \hat Z \hat P_G \hat Z^{-1}
+ 2 ( \gamma_\psi - \epsilon ) \hat Z \hat P_Q \hat Z^{-1}
\right\}_{jk}
\langle [ {\cal O}_{kl} ] \X \rangle \nonumber\\
&&\qquad+ i \frac{\beta_\epsilon}{\g}
\langle [ {\cal O}_{jl} \Delta^\g ] \X \rangle
+ i \sigma
\langle [ {\cal O}_{jl} \Delta^\xi ] \X \rangle
- i ( \gamma_\psi - \epsilon )
\langle [ {\cal O}_{jl} {\mit \Omega}_{\bar\psi\psi} ] \X \rangle
- i ( \gamma_G - \epsilon )
\langle [ {\cal O}_{jl} {\mit \Omega}_G ] \X \rangle ,
\end{eqnarray}
where
\begin{equation}
\hat\gamma_Z = - \mu \frac{d}{d \mu} \hat Z \, \hat Z^{-1}
= \g \frac{\partial}{\partial \g} \hat Z^{[1]}.
\end{equation}
Note, that only the whole anomalous dimension matrix of the composite
operators, $\hat\gamma \equiv \hat\gamma_Z + 2\gamma_G \hat P_G
+ 2 \gamma_\psi \hat P_Q$, is gauge independent. Consequently,
the off-diagonal elements of $\hat\gamma_Z$ are gauge independent too,
while its diagonal entries have a very simple dependence on the gauge
parameter.

Taking into account that the improved energy-momentum tensor
\cite{FreedWeinb74,Joglekar76,Nielsen77} in the massless gauge theory
admits the anomalous trace \cite{Collins77,Nielsen77,Minkowski76}
\begin{equation}
{\mit \Theta}_{\mu\mu} (x)
=
\frac{\beta_\epsilon}{\g}
[ \Delta^\g (x) ]
+ \sigma
[ \Delta^\xi (x) ]
- ( d - 2 ) [\Delta^{\rm ext} (x)]
- ( {\scriptstyle\frac{3}{2}} + \gamma_\psi - \epsilon )
{\mit \Omega}_{\bar\psi\psi} (x)
- ( 1 + \gamma_G - \epsilon ) {\mit \Omega}_G (x) ;
\end{equation}
and inserting it in the above equality (\ref{DWI}), making use of the
relation $ \langle [ {\cal O}_{jl} {\mit \Omega}_\phi ] \X \rangle
\equiv i \langle [ {\cal O}_{jl} ] N_\phi \{ \X \} \rangle$,
where $N_\phi$ is a number of $\phi$-particles operator, we come to
the familiar Callan-Symanzik equation \cite{CallanSymanzik} for the
Green function, albeit with a conformal operator insertion, where
the canonical scale dimensions of the field operators are shifted by the
anomalous ones \cite{Symanzik71}.

\subsection{Special conformal Ward identity.}

The derivation of the renormalized special conformal Ward identity
requires new inputs. We define the renormalization constants for the
product of the renormalized conformal operator and finite operator
insertions $[\Delta^i_-]$ as
\begin{equation}
i[{\cal O}_{jl}][\Delta^i_-]
=
i[{\cal O}_{jl} \Delta^i_-]
+ i \sum_{k = 0}^{j}
\{ \hat Z^\star_i \}_{jk}
[{\cal O}_{k l - 1}].
\end{equation}
For the product with EOM operator we have
\begin{equation}
i[{\cal O}_{jl}] {\mit \Omega}_\phi^-
=
i[{\cal O}_{jl} {\mit \Omega}_\phi^-]
- i \sum_{k = 0}^{j}
\{ \hat Z_\phi \}_{jk}
[{\cal O}_{k l - 1}].
\end{equation}
Simple calculations lead to the results
\begin{equation}
\label{Z-eff-const}
\hat Z_{\bar\psi\psi}
= 2 \hat Z \hat b (l) \hat P_Q \hat Z^{-1}, \quad
\hat Z_G
= 2 \hat Z \hat b (l) \hat P_G \hat Z^{-1} + \hat Z^\star_{\rm eff} ,
\end{equation}
where we have introduced a $\hat b$-matrix with elements
\begin{equation}
b_{jk}(l) = \theta_{jk} \{ 2(l+k+3)\delta_{jk}
- [1 + (-1)^{ j - k}] (2k + 3) \}.
\end{equation}
An additional renormalization constant in Eq. (\ref{Z-eff-const})
is extracted from the $\frac{1}{\epsilon}$-pole part of the operator
insertion
\begin{equation}
\label{Z-eff}
\int d^d x 2 x_-
\left\{
A_\rho (x) \frac{\delta}{\delta A_\rho (x)}
-
G_{+\rho} (x) \frac{\delta}{\delta G_{+\rho} (x)}
\right\} [{\cal O}_{jl}]
=
i \sum_{k = 0}^{j}
\{ \hat Z^\star_{\rm eff} \}_{jk} [{\cal O}_{kl-1}].
\end{equation}
In the non-local representation, to be used below, the left hand side of
this equality is given by
\begin{equation}
\label{W-NL}
\W ( \kappa_1, \kappa_2 )
=
\int d^d x 2 x_-
\left\{
A_\rho (x) \frac{\delta}{\delta A_\rho (x)}
-
G_{+\rho} (x) \frac{\delta}{\delta G_{+\rho} (x)}
\right\} {\cal O} (\kappa_1 , \kappa_2)
\end{equation}
in terms of the non-local string operators defined above.

Finally, the renormalized special conformal Ward identity takes the
form
\begin{eqnarray}
\label{SCWI}
&&\langle [ {\cal O}_{jl} ] \delta^C_- \X \rangle
= - i \sum_{k = 0}^{j}
\left\{ \hat a (l) + \hat \gamma^c (l) \right\}_{jk}
\langle [ {\cal O}_{kl-1} ] \X \rangle
+ i \frac{\beta_\epsilon}{\g}
\langle [ {\cal O}_{jl} \Delta^\g_- ] \X \rangle
+ i \sigma \langle [ {\cal O}_{jl} \Delta^\xi_- ] \X \rangle \\
&&\qquad - i (d - 2)
\langle [ {\cal O}_{jl} \Delta^{\rm ext}_- ] \X \rangle
- i ( \gamma_\psi - \epsilon )
\langle [ {\cal O}_{jl}
{\mit \Omega}_{\bar\psi\psi}^- ] \X \rangle
- i ( \gamma_G - \epsilon )
\langle [ {\cal O}_{jl} {\mit \Omega}_G^- ] \X \rangle .
\nonumber
\end{eqnarray}
Since we have expressed everything in terms of the renormalized
operator products, we can safely put the regularization parameter
$\epsilon = 0$. Moreover, because the LHS of this equality is finite
the same property is fulfilled by the RHS. We have defined above the
special conformal anomaly matrix
\begin{eqnarray}
\hat \gamma^c (l)
&=& \lim_{\epsilon \to 0} \Bigl\{
\hat Z \left[ \hat a (l)
- 2 ( \gamma_G - \epsilon ) \hat b (l) \hat P_G
- 2 ( \gamma_\psi - \epsilon ) \hat b (l) \hat P_Q
\right] \hat Z^{-1} \nonumber\\
&&\hspace{5.6cm}- ( \gamma_G - \epsilon ) \hat Z^\star_{\rm eff}
- \frac{\beta_\epsilon (\g)}{\g} \hat Z^\star_{\g}
- \gamma_\xi \hat Z^\star_{\xi} - \hat a (l) \Bigr\} \nonumber\\
&=& - 2 \left[ \gamma_G \hat P_G + \gamma_\psi \hat P_Q \right] \hat b
+ 2 [ \hat Z^{[1]} , \hat b ]_-
+ \hat Z^{\star[1]}_\g + \hat Z^{\star[1]}_{\rm eff},
\end{eqnarray}
where the second equality holds because $\hat\gamma^c$ is finite.

\section{Commutator constraint and two-loop singlet anomalous
dimensions.}

Now we are in a position to derive commutator constraint which induces
the off-diagonal elements of the anomalous dimension matrix. To this
end we will act with operator
\begin{eqnarray*}
D = \mu \frac{d}{d\mu} \hat 1 + \hat\gamma
\end{eqnarray*}
on both sides of the special conformal Ward identity (\ref{SCWI}).
This is completely equivalent to the use of the commutation relation
coming from the algebra of the collinear conformal group $\left[
\D, \K_- \right]_- = i \K_-$ and the combining use of the results
 (\ref{DWI}) and (\ref{SCWI}).

For this purpose we need the following commutators
\begin{eqnarray}
\label{g-commutator}
\left[
\mu \frac{d}{d \mu}, \g \frac{\partial}{\partial\g}
\right]_-
&=& - \left(
\g \frac{\partial}{\partial\g} \frac{\beta_\epsilon}{\g}
\right)
\g \frac{\partial}{\partial\g}
- \left(
\g \frac{\partial}{\partial\g} \sigma
\right)
\xi \frac{\partial}{\partial\xi} , \\
\label{xi-commutator}
\left[
\mu \frac{d}{d \mu}, \xi \frac{\partial}{\partial\xi}
\right]_-
&=& - \left(
\xi \frac{\partial}{\partial\xi} \sigma
\right)
\xi \frac{\partial}{\partial\xi} .
\end{eqnarray}
Since $\sigma = - 2\gamma_G$ is gauge independent in QED, the RHS of
Eq. (\ref{xi-commutator}) vanishes identically. One should be careful in
treatment of these equalities acting on the differential vertex operator
 insertions, since they are only valid  when both
differentials act on the same objects. Naive application could lead
to an error. An explicit derivation, which makes use of these
commutators,
gives the following renormalization group equations for the operator
insertions $[\Delta_-^i]$ in QED:
\begin{eqnarray}
\mu \frac{d}{d\mu} [\Delta_-^\g]
&=&
- \left(
\g \frac{\partial}{\partial\g} \frac{\beta}{\g}
\right)
[\Delta_-^\g]
- \left(
\g \frac{\partial}{\partial\g} \sigma
\right)
[\Delta_-^\g]
+ \left(
\g \frac{\partial}{\partial\g} \gamma_G
\right)
{\mit \Omega}_G^-
+ \left(
\g \frac{\partial}{\partial\g} \gamma_\psi
\right)
{\mit \Omega}_{\bar\psi\psi}^-
, \\
\mu \frac{d}{d\mu} [\Delta_-^\xi]
&=&
\left(
\xi \frac{\partial}{\partial\xi} \gamma_\psi
\right)
{\mit \Omega}_{\bar\psi\psi}^- .
\end{eqnarray}
The following steps are quite straightforward. Now we will act
repeatedly with the operator $D$ on every term in the special
conformal Ward identity without taking into account the action of
the latter on the field monomial $\X$, since at the end these terms
would anyway die out by the use of the conformal Ward identity.
First
\begin{eqnarray}
D \sum_{k = 0}^{j}
\left\{ \hat a (l) + \hat \gamma^c (l) \right\}_{jk}
\langle [ {\cal O}_{kl-1} ] \X \rangle
\!\!\!\!&=&\!\!\!\! \sum_{k = 0}^{j}
\left\{
\left[
\hat\gamma , \hat a + \hat\gamma^c + 2\frac{\beta}{\g} \hat b
\right]_- \right. \\
\!\!\!\!&-&\!\!\!\!
\left.
2 \beta \frac{\partial}{\partial\g}
\left[
\gamma_G \hat P_G
+
\gamma_\psi \hat P_Q
\right] \hat b
+
\beta \frac{\partial}{\partial\g}
\left[
\hat Z^{\star[1]}_\g
+
\hat Z^{\star[1]}_{\rm eff}
\right]
\right\}_{jk}
\langle [ {\cal O}_{kl-1} ] \X \rangle , \nonumber
\end{eqnarray}
where we have taken into account that (physical part of) the special
conformal anomaly matrix is gauge independent quantity. Next
\begin{eqnarray}
D \frac{\beta}{\g}
\langle [ {\cal O}_{jl} \Delta_-^\g ] \X \rangle
\!\!\!\!&=&\!\!\!\!
- \beta
\left( \frac{\partial}{\partial\g} \sigma \right)
\langle [ {\cal O}_{jl} \Delta_-^\xi ] \X \rangle
+ \beta
\left( \frac{\partial}{\partial\g} \gamma_G \right)
\langle [ {\cal O}_{jl} {\mit \Omega}_G^- ] \X \rangle
+ \beta
\left( \frac{\partial}{\partial\g} \gamma_\psi \right)
\langle [ {\cal O}_{jl} {\mit \Omega}_{\bar\psi\psi}^- ] \X \rangle
\nonumber \\
&+&\!\!\!\!\beta
\sum_{k = 0}^{j}
\left\{
\left(  \frac{\partial}{\partial\g} \hat Z^{\star[1]}_\g \right)
- 2 \frac{\partial}{\partial\g}
\left[
\gamma_G \hat P_G
+
\gamma_\psi \hat P_Q
\right] \hat b
\right\}_{jk}
\langle [ {\cal O}_{kl - 1} ] \X \rangle , \\
D \sigma
\langle [ {\cal O}_{jl} \Delta_-^\xi ] \X \rangle
\!\!\!\!&=&\!\!\!\!
\beta
\left( \frac{\partial}{\partial\g} \sigma \right)
\langle [ {\cal O}_{jl} \Delta_-^\xi ] \X \rangle
+ \sigma
\left( \xi\frac{\partial}{\partial\xi} \gamma_\psi \right)
\langle [ {\cal O}_{jl} {\mit \Omega}_{\bar\psi\psi}^- ] \X \rangle
\nonumber \\
&+&\!\!\!\!\sigma
\sum_{k = 0}^{j}
\left\{
\left(  \g\frac{\partial}{\partial\g} \hat Z^{\star[1]}_\xi \right)
- 2 \xi\frac{\partial}{\partial\xi}
\gamma_\psi \hat P_Q \hat b
\right\}_{jk}
\langle [ {\cal O}_{kl - 1} ] \X \rangle , \\
\label{RG-gluon-EOM}
D\gamma_G
\langle [ {\cal O}_{jl} {\mit \Omega}_G^- ] \X \rangle
\!\!\!\!&=&\!\!\!\!
\left(
\mu \frac{d}{d \mu} \gamma_G
\right)
\langle [ {\cal O}_{jl} {\mit \Omega}_G^- ] \X \rangle
-
\gamma_G
\sum_{k = 0}^{j}
\left\{
\g \frac{\partial}{\partial\g} \hat Z^{\star[1]}_{\rm eff}
\right\}_{jk}
\langle [ {\cal O}_{kl-1} ] \X \rangle , \\
D\gamma_\psi
\langle [ {\cal O}_{jl} {\mit \Omega}_{\bar\psi\psi}^- ] \X \rangle
\!\!\!\!&=&\!\!\!\!
\left(
\mu \frac{d}{d \mu} \gamma_\psi
\right)
\langle [ {\cal O}_{jl} {\mit \Omega}_{\bar\psi\psi}^- ] \X \rangle .
\end{eqnarray}
The result of Eq. (\ref{RG-gluon-EOM}) manifest the non-minimal
character of the subtraction, we have mentioned previously. Since the
LHS of each equation is finite we have omitted the divergent
contributions on the RHS. All other terms in the special conformal
Ward identity vanish by the action of the $D$-operator. Owing to
linear independence of the Gegenbauer polynomials we can safely omit
the Green function which enters in the special conformal Ward identity
and get the constraint
\begin{eqnarray}
\label{constraint}
\left[
\hat\gamma ,
\hat a (l) + \hat\gamma^c (l) + 2 \frac{\beta}{\g} \hat b (l)
\right]_- = 0 .
\end{eqnarray}
This equation shows that the off-diagonal part of $\hat\gamma$ in
$n$-loops can be obtained from the knowledge of the $(n-1)$-order of
the special conformal anomaly matrix $\hat\gamma^c$ and the
$\beta$-function. For instance, in the leading order we should keep
only the $\hat a$-matrix in Eq. (\ref{constraint}) as compared to the
last two entries since their perturbative expansion starts at
${\cal O} (\alpha)$. Therefore, from the diagonality of the
$\hat a (l)$-matrix it follows immediately that all entries in
$\hat\gamma^{(0)}$ (the first coefficient in the expansion $\hat\gamma
= \frac{\alpha}{2\pi} \hat\gamma^{(0)} + \left( \frac{\alpha}{2\pi}
\right)^2 \hat\gamma^{(1)} + \dots $) are diagonal too, and coincide
with the moments of the DGLAP evolution kernels.

In the NLO approximation the constraint (\ref{constraint}) fixes
unambiguously the off-diagonal elements of the anomalous dimension
matrices:
\begin{eqnarray}
{^{QQ}\!\gamma}_{jk}^{{\rm ND}(1)}
\!\!\!&=&\!\!\! \frac{1}{a(j,k)}
\left[
\left(
{^{QQ}\!\gamma}_{j}^{(0)} - {^{QQ}\!\gamma}_{k}^{(0)}
\right)
\left(
{^{QQ}\!\gamma}_{jk}^{c(0)} + \beta_0 b_{jk}
\right)
+
{^{QG}\!\gamma}_{j}^{(0)} {^{GQ}\!\gamma}_{jk}^{c(0)}
-
{^{QG}\!\gamma}_{jk}^{c(0)} {^{GQ}\!\gamma}_{k}^{(0)}
\right],
\\
{^{QG}\!\gamma}_{jk}^{{\rm ND}(1)}
\!\!\!&=&\!\!\! \frac{1}{a(j,k)}
\left[
{^{QQ}\!\gamma}_{j}^{(0)} {^{QG}\!\gamma_{jk}^{c(0)}}
-
{^{QQ}\!\gamma}_{jk}^{c(0)} {^{QG}\!\gamma}_{k}^{(0)}
\right],
\\
{^{GQ}\!\gamma}_{jk}^{{\rm ND}(1)}
\!\!\!&=&\!\!\! \frac{1}{a(j,k)}
\left[
{^{GQ}\!\gamma}_{j}^{(0)}
\left(
{^{QQ}\!\gamma}_{jk}^{c(0)} + \beta_0 b_{jk}
\right)
-
{^{GQ}\!\gamma}_{jk}^{c(0)} {^{QQ}\!\gamma}_{k}^{(0)}
+
{^{GG}\!\gamma}_{j}^{(0)} {^{GQ}\!\gamma}_{jk}^{c(0)}
\right],
\\
{^{GG}\!\gamma}_{jk}^{{\rm ND}(1)}
\!\!\!&=&\!\!\! \frac{1}{a(j,k)}
\left[
{^{GQ}\!\gamma}_{j}^{(0)} {^{QG}\!\gamma}_{jk}^{c(0)}
-
{^{GQ}\!\gamma}_{jk}^{c(0)} {^{QG}\!\gamma}_{k}^{(0)}
\right] .
\end{eqnarray}
Adding the two-loop results for the anomalous dimensions
$\gamma_j^{(1)}$ of the local operators without total
derivatives\footnote{Due to the particular normalization of
the Gegenbauer polynomials the $QG$ and $GQ$ results of Refs.
\cite{2-loop-DGLAP-PE,2-loop-DGLAP-PO} should be multiplied by
appropriate factors, i.e. $6/j$ and $j/6$, respectively.}, which
are available for about two decades from the calculations of
unpolarized \cite{2-loop-DGLAP-PE} and, recently, from polarized
\cite{2-loop-DGLAP-PO} deep inelastic scattering splitting
functions, we obtain the complete entry
\begin{equation}
\hat\gamma_{jk}^{(1)}
= \hat\gamma_j^{(1)} \delta_{jk}
+ \hat\gamma_{jk}^{{\rm ND}(1)} ,
\end{equation}
which governs the evolution of composite operators in Abelian gauge
theory to NLO.

\section{Evaluation of the conformal anomaly matrix.}

\begin{figure}[t]
\begin{center}
\vspace{4.7cm}
\hspace{-2cm}
\mbox{
\begin{picture}(0,220)(270,0)
\put(0,-30)                    {
\epsffile{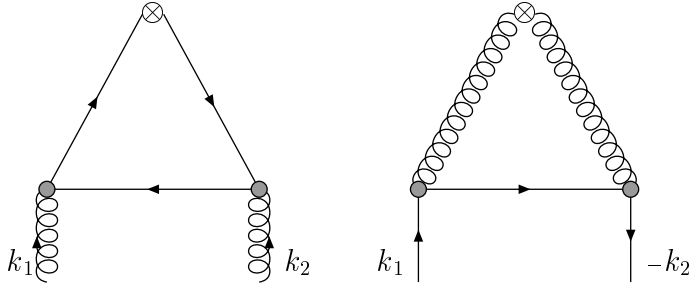}
                               }
\end{picture}
}
\end{center}
\vspace{-8.0cm}
\caption{\label{oneloop} Generic form of the one-loop diagrams for
the renormalization constants $Z^{\star}_i$. Grey blobs correspond
to the usual (Eq. (\protect\ref{usualFR})) or modified (Eq.
(\protect\ref{modifiedFR})) Feynman rules depending on the quantity
in question. The crossed circle stands for the usual non-local string
operator or for the modified operator insertion defined by Eq.
(\protect\ref{W-NL}) which determines the $\hat w$-matrix.}
\end{figure}

\subsection{Local anomalous dimensions.}

In this section we evaluate all one-loop renormalization constants
which enter in the special conformal anomaly matrix and determine the
anomalous dimensions ${^{AB}\!\gamma}_{jk}^{{\rm ND}(1)}$.

First, in the leading order $2 \hat Z^{[1]} = \frac{\alpha}{2\pi} \hat
\gamma^{(0)}_Z $. Second, the calculation of the diagrams depicted in
Fig. \ref{oneloop} and the extraction of the pole part from the
renormalization constant $\hat Z^\star_\g = \frac{1}{\epsilon}
\hat Z^{\star [1]}_\g +\cdots $ gives:
\begin{equation}
{^{AB}\!{\hat Z}}^{\star [1]}_\g
= - \frac{\alpha}{2\pi} {^{AB}\!{\hat \gamma}}^{(0)}_Z \hat b ,
\quad \mbox{for\ } A \neq B.
\end{equation}
Here we took into account that according to the Feynman rules
(\ref{modifiedFR}) these diagrams are given by the derivatives with
respect to the external momenta of the familiar one-loop graphs; and
that the differentiation of the Gegenbauer polynomials leads
to the $\hat b$-matrix:
\begin{equation}
\label{b-diff}
\left(
\frac{\partial}{\partial k_{1+}} + \frac{\partial}{\partial k_{2+}}
\right)
(k_{1+} + k_{2+})^{l - 1} C^\nu_j
\left(
\frac{k_{1+} - k_{2+}}{k_{1+} + k_{2+}}
\right)
=
\sum_{k = 0}^{j} b_{jk} (l)
(k_{1+} + k_{2+})^{l - 2} C^\nu_j
\left(
\frac{k_{1+} - k_{2+}}{k_{1+} + k_{2+}}
\right) .
\end{equation}
Defining also the one-loop approximation of the constant
$\hat Z^\star_{\rm eff} = \frac{1}{\epsilon} \hat
Z^{\star [1]}_{\rm eff} +\cdots  $ in terms of the $\hat w$-matrix,
\begin{eqnarray}
{^{AB}\hat Z^{\star [1]}_{\rm eff}}
&=& \frac{\alpha}{2\pi} {^{AB} \hat w},
\end{eqnarray}
we finally obtain the result for the special conformal anomaly to
leading order:
\begin{equation}
{^{AB}\!\gamma}_{jk}^{c(0)}
= -b_{jk}{^{AB}\!\gamma}_{k}^{(0)} + {^{AB}\!w}_{jk} .
\end{equation}
The first entries, which are the usual anomalous dimensions
of conformal operators, are well known and read
\begin{eqnarray}
\label{anomalous-dimensions}
{^{QQ}\!\gamma}_{j}^{(0)}
&=&
- C_F \left(
3 + \frac{2}{( j + 1 )( j + 2 )} - 4 \psi( j + 2 ) + 4 \psi(1)
\right)
\\
{^{QG}\!\gamma}_{j}^{(0)}
&=&
\frac{-24 N_f T_F}{j( j + 1 )( j + 2 )( j + 3 )}
\times \left\{
j^2 + 3 j + 4,\quad \mbox{for even parity}
\atop
j( j + 3 ),\quad\quad  \mbox{for odd parity}
\right.
\\
{^{GQ}\!\gamma}_{j}^{(0)}
&=&
\frac{-C_F}{3( j + 1 )( j + 2 )}
\times \left\{
j^2 + 3 j + 4 ,\quad \mbox{for even parity}
\atop
j ( j + 3 ),\quad\quad \mbox{for odd parity}
\right.
\\
\label{last-AD}
{^{GG}\!\gamma}_{j}^{(0)}
&=& 2 \gamma_G^{(0)}
= \frac{4}{3} T_F N_f .
\end{eqnarray}
The calculation of the $\hat w$-matrix for the $QQ$-sector can be found
in Ref. \cite{Mul94}. For the $GQ$-channel we employ the non-local
representation for the evaluation of the corresponding kernel. Simple
calculation provides the result
\begin{equation}
\W^{\mit\Gamma} ( \kappa_1 , \kappa_2 )
= \frac{i}{\epsilon} \frac{\alpha}{2\pi}
\int_{0}^{1} dy \int_{0}^{\bar y} dz
{^{GQ}{\cal K}^w} (y, z)
{^Q\!{\cal O}^{\mit\Gamma}}
(\bar y \kappa_1 + y \kappa_2 , z \kappa_1 + \bar z \kappa_2 ),
\end{equation}
with the same kernel in the parity even and odd channels
\begin{equation}
{^{GQ}{\cal K}^w} (y, z) =
C_F [ \delta (z) - \delta (y) ] .
\end{equation}
Note that this kernel is gauge independent as it should be (the
$\xi$-dependent piece we have found in the calculation is
symmetrical with respect to the permutation $y \leftrightarrow z$,
and vanishes because of the symmetry properties of the operators).
The evaluation of the conformal moments of this simple kernel is
quite straightforward and we get
\begin{eqnarray}
{^{QQ}\!w}_{jk}
&=&
- C_F \theta_{j-2,k}\left[ 1 + (-1)^{j-k} \right]
\frac{( 2 k + 3 ) a(j,k)}{( k + 1 ) ( k + 2 )}
\nonumber\\
&&\hspace{4cm}
\times\left[
A_{jk} - \psi( j + 2 ) + \psi(1)
+ 4 ( k + 1 ) ( k + 2 ) \frac{A_{jk}}{a(j,k)}
\right] ,
\\
\mbox{with}\!\!\!\!\!\!\!\!\!\!&& A_{jk} =
\psi\left( \frac{j + k + 4}{2} \right)
-\psi\left( \frac{j-k}{2} \right)
+ 2 \psi ( j - k ) - \psi ( j + 2 ) - \psi(1),
\nonumber
\\
\label{w-GQ}
{^{GQ}\!w}_{jk}
&=&
- C_F \theta_{j-2,k} \left[ 1 + (-1)^{j-k} \right]
\frac{1}{6}
\frac{( 2 k + 3 ) a(j,k)}{( k + 1 ) ( k + 2 )} .
\end{eqnarray}
In addition ${^{QG}\!w} = {^{GG}\!w} = 0$.
Defining new matrices $d_{jk} = b_{jk}/a(j,k)$ and $g_{jk}
= w_{jk}/a(j,k)$, we can rewrite the off-diagonal elements:
\begin{eqnarray}
\label{andimND-QQ}
{^{QQ}\!\gamma}_{jk}^{{\rm ND}(1)}
&=&
\left(
{^{QQ}\!\gamma}_{j}^{(0)} - {^{QQ}\!\gamma}_{k}^{(0)}
\right)
\left(
d_{jk}
\left\{
\beta_0 - {^{QQ}\!\gamma}_{k}^{(0)}
\right\}
+ {^{QQ}\!g}_{jk}
\right)
-
\left(
{^{QG}\!\gamma}_{j}^{(0)} - {^{QG}\!\gamma}_{k}^{(0)}
\right) d_{jk}
{^{GQ}\!\gamma}_{k}^{(0)} \nonumber\\
&&\hspace{9cm}
+ {^{QG}\!\gamma}_{j}^{(0)} {^{GQ}\!g}_{jk},
\\
\label{andimND-QG}
{^{QG}\!\gamma}_{jk}^{{\rm ND}(1)}
&=&
- \left(
{^{QQ}\!\gamma}_{j}^{(0)} - {^{QQ}\!\gamma}_{k}^{(0)}
\right)
d_{jk} {^{QG}\!\gamma}_{k}^{(0)}
-
{^{QQ}\!g}_{jk} {^{QG}\!\gamma}_{k}^{(0)} ,
\\
\label{andimND-GQ}
{^{GQ}\!\gamma}_{jk}^{{\rm ND}(1)}
&=&
\left(
{^{GQ}\!\gamma}_{j}^{(0)} - {^{GQ}\!\gamma}_{k}^{(0)}
\right) d_{jk}
\left\{
\beta_0 - {^{QQ}\!\gamma}_{k}^{(0)}
\right\}
+
{^{GQ}\!\gamma}_{j}^{(0)} {^{QQ}\!g}_{jk}
+
{^{GQ}\!g}_{jk}
\left\{
\beta_0 - {^{QQ}\!\gamma}_{k}^{(0)}
\right\},
\\
\label{andimND-GG}
{^{GG}\!\gamma}_{jk}^{{\rm ND}(1)}
&=&
- \left(
{^{GQ}\!\gamma}_{j}^{(0)} - {^{GQ}\!\gamma}_{k}^{(0)}
\right) d_{jk}
{^{QG}\!\gamma}_{k}^{(0)}
-
{^{GQ}\!g}_{jk}{^{QG}\!\gamma}_{k}^{(0)}.
\end{eqnarray}

\subsection{Evolution kernels.}

The representations (\ref{andimND-QQ}-\ref{andimND-GG}) allow us to
write the off-diagonal part of the evolution kernels as convolution
of known kernels. The LO kernels ${^{AB}\!V^{(0)}}$ in the expansion
$V ( x, y | \alpha) = \frac{\alpha}{2\pi} V^{(0)} ( x, y )
+ \left( \frac{\alpha}{2\pi} \right)^2 V^{(1)}( x, y ) + \dots$ are
diagonal with respect to Gegenbauer polynomials and their eigenvalues
are $-{^{AB}\!\gamma}_j^{(0)}/2$:
\begin{equation}
\int_{0}^{1} dx
\Biggl(\!
\begin{array}{c}
C^{\frac{3}{2}}_j \\
C^{\frac{5}{2}}_{j - 1}
\end{array}
\!\Biggr) (2x - 1)
\Biggl({
{^{QQ} V^{(0)}}\ {^{QG} V^{(0)}}\atop
{^{GQ} V^{(0)}}\ {^{GG} V^{(0)}}
}\Biggr) (x, y)
=
-\frac{1}{2}
\Biggl({
{^{QQ}\!\gamma_j^{(0)}}\ {^{QG}\!\gamma_j^{(0)}}\atop
{^{GQ}\!\gamma_j^{(0)}}\ {^{GG}\!\gamma_j^{(0)}}
}\Biggr)
\Biggl(\!
\begin{array}{c}
C^{\frac{3}{2}}_j \\
C^{\frac{5}{2}}_{j - 1}
\end{array}
\!\Biggr) (2y - 1).
\end{equation}
These results are already known or can be derived from the DGLAP
kernels by the  procedure described below. In the restricted region
$|x,y|\le1$ these kernels have the general structure
\begin{equation}
{^{AB} V}^{(0)}( x, y )
=
\theta( y - x )
\, {^{AB}\!F} ( x, y ) \pm
\left\{ x \to \bar x \atop  y \to \bar y \right\},
\mbox{\ for\ }
\left\{ {\mbox{$A=B$} \atop \mbox{$A \not= B$}}\right. ,
\end{equation}
with
\begin{eqnarray}
{^{QQ}\!F} ( x, y )
&=& C_F \frac{x}{y}
\left[ 1 + \frac{1}{(y-x)_+} + \frac{3}{2} \delta(x-y) \right] ,
\\
\label{V-QG}
{^{QG}\!F} ( x, y )
&=&
2 N_f T_F \frac{x}{y^2 \bar y} \left\{
2 x - y - 1 , \quad \mbox{\ for even parity} \atop
- \bar y , \qquad\qquad \mbox{\ for odd parity} \right. ,
\\
\label{V-GQ}
{^{GQ}\!F} ( x, y )
&=& C_F \frac{x}{y} \left\{
2 y - x , \quad \mbox{\ for even parity} \atop
x , \qquad\quad \mbox{\ for odd parity} \right. ,
\\
{^{GG}\!F} ( x, y )
&=& - \frac{\beta_0}{2} \delta(x-y).
\end{eqnarray}
Here one comment is in order to clarify the differences of the
even parity $QG$ and $GQ$ kernels available in the literature.
Indeed these kernels are not uniquely defined and different
expressions could be obtained from various calculation techniques
\cite{Cha80,Ohr80,Radyushkin97,BGR97,Ji97}. However, apparently
different functional dependence on the momentum fraction $x,y$ leads
to the same eigenvalues in the physical sectors of the corresponding
channels. The only difference arises in the space of the
non-physical moments. Namely, the kernels derived by Chase \cite{Cha80}
(and listed above) possess the off-diagonal matrix elements in the basis
of the Gegenbauer polynomials in the non-physical sector, while the
kernels derived below
\begin{equation}
\label{improved}
{^{QG}\!F}_{\rm impr} ( x, y )
= 2 N_f T_F \frac{x}{y^2} ( - 1 + 2 x - 4 \bar x y ) ,
\quad
{^{GQ}\!F}_{\rm impr} ( x, y )
= C_F \frac{x^2}{y} ( - 1 + 2 y + 4 \bar x y )
\end{equation}
are exactly diagonal.
The difference can be attributed to the following additional pieces
$\Delta {^{QG}\!F} ( x, y )$ $= 4 N_f T_F \frac{x \bar x}{y \bar y}
(2y - 1)$ and $\Delta {^{GQ}\!F} ( x, y ) = 2 C_F x \bar x (2x - 1)$
which should be added to the results of Chase to get the equations
given behind. Note also that the kernels we have listed (taken without
colour factors) respect the SUSY relation \cite{BFKL85}:
\begin{equation}
{^{GQ} V}^{(0)}( x, y ) y \bar y
=
{^{QG} V}^{(0)}( y, x ) (x \bar x)^2 .
\end{equation}

Now we are able to reconstruct the off-diagonal parts of the singlet
two-loop ER-BL kernels from the local anomalous dimensions we have
derived above, i.e. $g_{jk}$ and $( {^{AB}\!\gamma}_j^{(0)} -
{^{AB}\!\gamma}_k^{(0)} )d_{jk}$ matrices.
In the $QQ$-channel the $g$-kernel is given in Ref. \cite{Mul94}, while
for the $GQ$-channel we have been able to perform the summation of
the Gegenbauer polynomials by the procedure discussed below in
section~\ref{ER-BLtoDGLAP}:
\begin{eqnarray}
\label{g-kerQQ}
{^{QQ}\!g}(x,y)
&=& - C_F
\left[ \theta( y - x )
\frac{ \ln \left( 1 - \frac{x}{y} \right) }{y - x} +
\left\{x\to \bar{x} \atop y\to \bar{y} \right\} \right]_+,
\\
\label{g-kerGQ}
{^{GQ}\!g} ( x, y )
&=& - C_F
\left[
\theta(y-x) \frac{x}{y} ( 2 x - 1 ) -
\left\{ x \to \bar x \atop y \to \bar y \right\}
\right] .
\end{eqnarray}
Intuitively, one could anticipate this last result. For that one should
note that in Eq. (\ref{w-GQ}) the factor $1/(k+1)(k+2)$ gives just the
eigenvalues of the  ER-BL kernel in scalar $\phi^3_{(6)}$-theory. Next,
the off-diagonal matrix elements of the $\hat b$-matrix are $(2k + 3)$.
According to Eq. (\ref{b-diff}), the latter can be induced by derivative
of the Gegenbauer polynomials $C_k^\nu (2x - 1)$ with respect to the
argument $x$ times the factor $(2x - 1)$. Since the derivative increase
the index and reduce the order by one unity, we come to the equation we
have just presented.

Note that the kernels (\ref{g-kerQQ},\ref{g-kerGQ}) possess also
diagonal conformal moments. They can be subtracted by applying the
projection operator $( \I - \D )$, where $\D$ extracts the diagonal
part of any test function $\tau (x, y)$:
\begin{eqnarray*}
\int_{0}^{1} dx C^\nu_j (2x - 1)
\left\{
\begin{array}{c}
\I \\
\D
\end{array}
\right\}
\tau (x, y)
=
\sum_{ k = 0 }^{j}
\tau_{jk}
\left\{
\begin{array}{c}
1 \\
\delta_{jk}
\end{array}
\right\}
C^\nu_k (2y - 1).
\end{eqnarray*}
The kernel ${^{AB}\!g}(x,y)$ that possesses the
${^{AB}\!g}_{ij}$-moments is defined as follows
\begin{equation}
\int_{0}^{1} dx
\Biggl(\!
\begin{array}{c}
C^{\frac{3}{2}}_j \\
C^{\frac{5}{2}}_{j - 1}
\end{array}
\!\Biggr) (2x - 1)
( \I - \D ) {^{AB}\!g} ( x, y )
=
\sum_{k = 0}^{j-2} {^{AB}\!g}_{jk}
\Biggl(\!
\begin{array}{c}
C^{\frac{3}{2}}_k \\
C^{\frac{5}{2}}_{k - 1}
\end{array}
\!\Biggr) (2y - 1).
\end{equation}

The factor $({^{AB}\!\gamma}_j^{(0)}-{^{AB}\!\gamma}_k^{(0)}) d_{jk}$
corresponds to the expected breaking of the diagonality of the NLO
exclusive kernels \cite{BDFL86,MikhRad86} and it is related to the
one-loop renormalization of the composite operators. These contributions
can be embedded into a redefined conformal representation, which differs
from the original one by the shift of the scale dimensions of local
operators by the anomalous ones $\gamma_j$. This means that new
eigenfunctions should be orthogonal to each other with the weight
factor $(y \bar y)^{n + \gamma_j}$, with $n=1,2$ corresponding to
quarks and gluons, respectively. This gives us a guide how to construct
new kernels $V_\delta (x, y)$ from the leading order ones. For that
we take into account that $V_\delta (x, y)$ should be symmetrical
after multiplication by the factor $(y \bar y)^{n + \delta}$:
\begin{equation}
{^{AA} V}_\delta (x, y) (y \bar y)^{n + \delta}
=
{^{AA} V}_\delta (y, x) (x \bar x)^{n + \delta} ,
\end{equation}
with $n=1,2$ for $A=Q,G$-channels, respectively, and
\begin{equation}
{^{GQ} V}_\delta ( x, y ) ( y \bar y )^{1 + \delta}
=
{^{QG} V}_\delta ( y, x ) (x \bar x)^{2 + \delta} ,
\end{equation}
for mixed ones. These equations tell us that the two-loop off-diagonal
terms $\dot{V} ( x, y )$ should be proportional to the first term
in the Taylor expansion of $V_\delta (y, x)$ with respect to $\delta$.
This recipe leads to the following dotted kernels
\begin{eqnarray}
\label{dot-ker}
{^{AB} \dot{V}} ( x, y )
&=& \theta(y-x) \,
\left[
{^{AB}\!F}(x,y) \ln\frac{x}{y}
+
\Delta {^{AB}\!\dot F} ( x, y )
\right]
\pm
\left\{ x \to \bar x \atop y \to \bar y \right\},
\mbox{\ for\ } \left\{ {\mbox{$A=B$} \atop \mbox{$A\not=B$}} \right. .
\end{eqnarray}
Note, however, that simple logarithmic modification of the leading
order result does not work for the parity even mixing channels. The
main difference from the other cases comes from the fact that Chase's
kernels possess off-diagonal elements in the non-physical space of
conformal moments, which immediately manifest themselves when these
kernels are multiplied by factor $\ln(x/y)$. However, the above symmetry
consideration could not fix unambiguously the modification of extra
pieces $\Delta F$ and, therefore, it requires an explicit analysis. We
find the following results for $\Delta {^{AB}\!\dot F}$:
\begin{equation}
\label{AddKer}
\Delta {^{GQ}\!{\dot F}^V}
= 2 C_F
\left[
x \ln y - \bar x \ln \bar x
\right],
\quad
\Delta {^{QG}\!{\dot F}^V}
= 4 T_F N_f \frac{1}{y \bar y}
\left[
x \ln y - \bar x \ln \bar x
\right] .
\end{equation}

Combining the equations
(\ref{g-kerQQ},\ref{g-kerGQ},\ref{dot-ker},\ref{AddKer}),
the off-diagonal parts of the NLO evolution kernels read:
\begin{eqnarray}
{^{QQ} V}^{{\rm ND}(1)}
&=& - ( \I - \D )
\biggl(
{^{QQ}\dot{V}}
\otimes
\left\{
{^{QQ} V}^{(0)} + \frac{\beta_0}{2} I
\right\}
+ {^{QQ}\!g}
\otimes {^{QQ} V}^{(0)}
-
{^{QQ} V}^{(0)}  \otimes {^{QQ}\!g} \nonumber\\
&&\hspace{7cm} +
{^{QG}\dot{V}} \otimes {^{GQ}V}^{(0)}
-
{^{QG} V}^{(0)} \otimes {^{GQ}\!g} \biggr),
\\
{^{QG}V}^{{\rm ND}(1)}
&=& - ( \I - \D )
\left(
{^{QQ}\dot{V}} \otimes {^{QG} V}^{(0)} +
{^{QQ}\!g} \otimes {^{QG}V}^{(0)}
\right)
\\
{^{GQ} V}^{{\rm ND}(1)}
&=& - ( \I - \D )
\biggl(
{^{GQ}\dot{V}}
\otimes
\left\{
{^{QQ} V}^{(0)} + \frac{\beta_0}{2} I
\right\}
-
{^{GQ}V}^{(0)} \otimes {^{QQ}\!g} \nonumber\\
&&\hspace{7cm}+
{^{GQ}\!g}
\otimes
\left\{
{^{QQ}V}^{(0)} + \frac{\beta_0}{2} I
\right\}
\biggr) ,
\\
{^{GG} V}^{{\rm ND}(1)}
&=& - ( \I - \D )
\left(
{^{GQ}\dot{V}} \otimes {^{QG}V}^{(0)}
+
{^{GQ}\!g}
\otimes
{^{QG}V}^{(0)}
\right).
\end{eqnarray}
Here we use a shorthand $\otimes = \int_{0}^{1} dz$.

\section{Fermion bubble insertions in the singlet kernels.}

\begin{figure}[t]
\begin{center}
\vspace{4.7cm}
\hspace{-2cm}
\mbox{
\begin{picture}(0,220)(270,0)
\put(0,-30)                    {
\epsffile{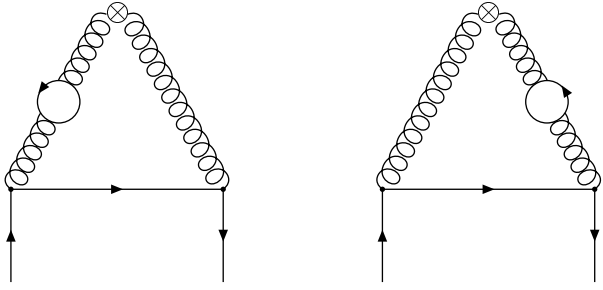}
                               }
\end{picture}
}
\end{center}
\vspace{-8.0cm}
\caption{\label{bubble} Diagrams giving rise to the $\beta_0$
dependence in the NLO evolution $GQ$-kernel.}
\end{figure}

In this section we present the check of the self-consistency of
our approach by explicit evaluation of the fermion vacuum insertions
in the $GQ$-evolution kernels. From the conformal constraints
we know that corresponding contribution to the off-diagonal part
of the evolution kernel is of the form
\begin{equation}
\Delta {^{GQ} V}^{{\rm ND}(1)}
= - \frac{\beta_0}{2}
( \I - \D )
\biggl(
{^{GQ}\dot{V}} ( x, y ) + {^{GQ}\!g} ( x, y )
\biggr).
\end{equation}
Therefore, evaluating the diagrams in Fig. \ref{bubble} we should
immediately obtain the desired form of the dotted and $g$-kernels. To
this end we have exploited the formalism developed in Refs.
\cite{MikhRad85,MikhRad86}, which is the most appropriate for the
multi-loop calculations of the evolution kernels and makes use of
the, so-called, $\alpha$-representation for the Feynman propagators.
Corresponding calculations run along the same line as for the
one-loop diagrams but now we insert the fermion vacuum polarization
$i \Pi_{\mu\nu} (k)$ into the gluon lines (see Fig. \ref{bubble})
\begin{equation}
\Pi_{\mu\nu} (k) = \Pi (\epsilon)
\left(
\frac{\mu^{*2}}{-k^2}
\right)^\epsilon \left( k^2 g_{\mu\nu} - k_\mu k_\nu \right) ,
\quad\mbox{with}\quad
\Pi (\epsilon) =
\frac{1}{\epsilon}
\frac{\alpha}{4\pi} (-\beta_0)
\frac{6 \Gamma (1 + \epsilon)
\Gamma^2 (2 - \epsilon) }{\Gamma (4 - 2\epsilon)},
\end{equation}
(with $\mu^{*2} = 4\pi \mu^2$), which results to the mere shift of
the power of the denominator of the gluon propagator by the magnitude
$\epsilon$. Since we are interested only in the off-diagonal part
of the corresponding diagrams we can safely omit, after subtraction
of sub-divergences, the contribution coming from the surviving
overall constants, i.e. $\Pi(\epsilon)$ etc., since it would be
proportional to the leading order evolution kernel. The remainder
of the graphs leads to the integrals
\begin{equation}
\label{bubble-integral}
I = \left. \frac{\partial}{\partial\epsilon} \right|_{\epsilon = 0}
\int_{0}^{1} d\alpha_1 \int_{0}^{\bar \alpha_1} d\alpha_2
\left[ \alpha_1^\epsilon + \alpha_2^\epsilon \right]
\left\{
v_1 (\alpha_1,\alpha_2)
+
v_2 (\alpha_1,\alpha_2)
\frac{\partial}{\partial x}
\right\}
\delta (x - y \bar \alpha_1 - \bar y \alpha_2),
\end{equation}
where the functions $v_i$ read for the parity even and odd channels
\begin{eqnarray}
v_1^V (\alpha_1,\alpha_2)
&=& - \bar \alpha_2 + y ( \bar \alpha_2 + \bar \alpha_1 ) ,
\\
v_1^A (\alpha_1,\alpha_2)
&=& - \bar \alpha_2 + \alpha_2 + 2 y ( \bar \alpha_2 - \alpha_1 ) ,
\\
v_2^V (\alpha_1,\alpha_2) &=& v_2^A (\alpha_1,\alpha_2) \nonumber\\
&=& \alpha_2 \bar \alpha_2
+ y
\left(
\bar \alpha_1 + \alpha_2 - 2 \alpha_2 (\bar \alpha_1 + \bar \alpha_2)
\right)
- y^2
\left(
\bar \alpha_1^2 + \alpha_2^2 - 2 \bar \alpha_1 \alpha_2
\right) .
\end{eqnarray}
To perform the final integrations we need the following formulae
\cite{BM97}:
\begin{equation}
\int_{0}^{1} d \alpha \delta (x - y \alpha) = \Theta_{11}^0 (x, x - y),
\end{equation}
for the first one, and
\begin{equation}
\int_{0}^{1} d \alpha \bar \alpha \alpha^\tau
\Theta_{11}^0 (x - y \bar \alpha , \alpha - \bar x )
=
\frac{1}{\tau + 1}
\left\{
\frac{1}{\bar y}
\left[ \bar x^{\tau + 1}
- \left( 1 - \frac{x}{y} \right)^{\tau + 1} \right]
\theta (y - x)
+
\frac{ \bar x^{\tau + 1}}{\bar y} \theta (x - y)
\right\} ,
\end{equation}
for the second. Putting $\epsilon = 0$ in the square brackets in
Eq. (\ref{bubble-integral}), we immediately obtain twice the leading
order kernels\footnote{To be precise, in the parity even channel,
we get Chase's kernel.}. Evaluating the integral for arbitrary
$\epsilon$ and differentiating it with respect to $\epsilon$, we
finally obtain the contribution to the two-loop kernel
\begin{eqnarray}
&&\Delta {^{GQ} V^{(1)}} (x, y)
= - \frac{\beta_0}{2}
\biggl\{
F (x, y)
+
{^{GQ}\!F} (x, y) \ln x
-
{^{GQ}\!F} (\bar x, \bar y) \ln \bar x \\
&&\hspace{5cm}+
\left[
{^{GQ}\!F} (x, y) + {^{GQ}\!F} (\bar x, \bar y)
\right]
\ln \left( 1 - \frac{x}{y} \right)
\biggr\} \theta (y - x)
-
\left\{
{ x \to \bar x \atop y \to \bar y }
\right\} , \nonumber
\end{eqnarray}
with
\begin{equation}
F (x, y) =
2 C_F \frac{x}{y}
\left\{
1 - 2 y , \quad \mbox{\ for even parity} \atop
1 - 2 x , \quad \mbox{\ for odd parity}
\right. ,
\end{equation}
while the other functions are given by Eqs. (\ref{V-QG},\ref{V-GQ}).
Obviously, these expressions consist of both diagonal and off-diagonal
parts of the evolution kernel. Subtracting the sum of the dotted and
the $g$-kernels derived in the preceding section it is a trivial task
to convince oneself that the remainder is diagonal in the basis of the
Gegenbauer polynomials. Namely, this provides a new diagonal element
\begin{eqnarray}
U^{\rm D} ( x , y )
=
\biggl\{
{^{GQ}\!g} (x, y)
&+&
{^{GQ}\!F^A} (x, y) \ln y
-
{^{GQ}\!F^A} (\bar x, \bar y) \ln \bar x \\
&+&
\left[
{^{GQ}\!F^A} (x, y) + {^{GQ}\!F^A} (\bar x, \bar y)
\right]
\ln \left( 1 - \frac{x}{y} \right)
\biggr\} \theta (y - x)
-
\left\{
{ x \to \bar x \atop y \to \bar y }
\right\} , \nonumber
\end{eqnarray}
and an equivalent representation for the $g$-kernel in the
$GQ$-channel (\ref{g-kerGQ})
\begin{equation}
{^{GQ}\!g}(x,y)
= - C_F
\left[
\theta( y - x )
\ln \left( 1 - \frac{x}{y} \right) -
\left\{x\to \bar{x} \atop y\to \bar{y} \right\}
\right] ,
\end{equation}
which holds true for the off-diagonal matrix elements.
 Note that it is most close to the functional
form of ${^{QQ}\!g}$.

Finally, we can conclude that an explicit calculation supports our
results derived from the use of the conformal consistency relation
(\ref{constraint}).

\section{DGLAP $\to$ ER-BL reduction.}
\label{ER-BLtoDGLAP}

Aiming an explicit numerical integration of the two-loop evolution
equations one is interested in the momentum fraction splitting kernels.
While the off-diagonal parts of these kernels are known explicitly, it
is not the case for the diagonal entries. In this section we outline the
formalism which fills this gap. We have been able to restore the one-loop
ER-BL evolution kernels analytically. The reduction formulae we have
derived here are sufficient to find numerically the diagonal part (in the
basis of the Gegenbauer polynomials) of the exclusive singlet kernels
from the corresponding DGLAP analogues.

\subsection{$QQ$ channel.}


\begin{figure}[t]
\begin{center}
\vspace{3cm}
\hspace{-8.3cm}
\mbox{
\begin{picture}(0,220)(270,0)
\put(0,-30)                    {
\epsffile{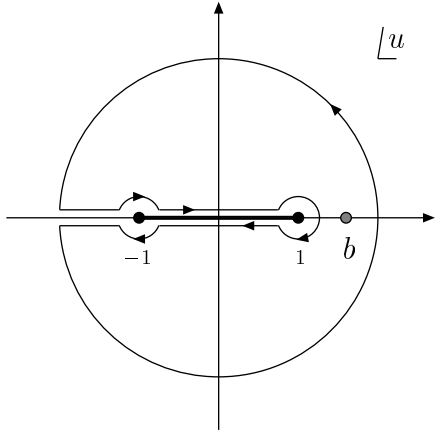}
                               }
\end{picture}
}
\end{center}
\vspace{-5.0cm}
\caption{\label{contour} The integration contour for the evaluation
of the integral (\protect\ref{integral}). There is a $n$-th order
pole in the point $u = b$ and the cut on the real axis from $-1$ to
$1$.}
\end{figure}

The formal expression for the diagonal part of the ER-BL evolution
kernel can be given by the following formula ($w(x | \nu) = (x \bar
x)^{\nu - \frac{1}{2}}$)
\begin{equation}
V^{\rm D} ( x, y ) = - \frac{1}{2}
\sum_{j = 0}^{\infty} \frac{w(x | \nu)}{N_j(\nu)}
C^\nu_j (2x - 1) \gamma^j C^\nu_j (2y - 1),
\end{equation}
where $N_j(\nu)= 2^{ - 4 \nu + 1 }
\frac{ \Gamma^2 (\frac{1}{2}) \Gamma ( 2 \nu + j )}
{\Gamma^2 (\nu) ( \nu + j ) j! }$ is a normalization factor and
$\gamma_j$ are the anomalous dimensions of the conformal operators
(see, for instance, Eqs. (\ref{anomalous-dimensions}-\ref{last-AD})).

Therefore, the main problem is to perform the infinite summation and
construct the generating function for the exclusive evolution kernels.
Using the Gegenbauer's summation theorem for $C^\nu_j$-polynomials
\cite{BE53_1} this goal can be easily achieved. We have evaluated the
following conversion function from the forward to non-forward limits:
\begin{eqnarray}
\label{DGLAP->ER-BL}
&&\sum_{j = 0}^{\infty} \frac{w(x | \nu)}{N_j(\nu)}
C^\nu_j (2x - 1) z^j C^\nu_j (2y - 1)
=
2^{4\nu - 1}
\frac{\Gamma(\nu)\Gamma(\nu + 1)}{\Gamma^2(\frac{1}{2})\Gamma(2\nu)}
(x \bar x)^{\nu - \frac{1}{2}}(1 - z^2)
\\
&&\qquad\qquad\qquad\times
\left[
1 - 2 \left( (2x - 1)(2y - 1)
- 4 \sqrt{x \bar x y \bar y} \right) z + z^2
\right]^{-\nu - 1}
\nonumber\\
&&\qquad\qquad\qquad\qquad\qquad\qquad\times
{_2F_1}
\left( { \nu + 1, \nu \atop 2 \nu }
\left| - \frac{16 \sqrt{x \bar x y \bar y} z}{1
- 2 \left( (2x - 1)(2y - 1) - 4 \sqrt{x \bar x y \bar y} \right) z + z^2 }
\right.\right).
\nonumber
\end{eqnarray}
In order to find the ER-BL evolution kernel we have to convolute
the above equality with the DGLAP splitting function and evaluate the
integral on the RHS of this equation. For this purpose we use
the Euler's integral representation of the hypergeometric function,
so that the above formula looks more compact (here $a (u) =
(2x - 1)(2y - 1) + 4 u \sqrt{x \bar x y \bar y}$):
\begin{equation}
\sum_{j = 0}^{\infty}
\frac{w(x | \nu)}{N_j(\nu)}
C^\nu_j (2x - 1) z^j C^{\nu}_j (2y - 1)
= 2^{2\nu} \frac{\nu}{\pi} (x \bar x)^{\nu - \frac{1}{2}}
\int_{-1}^{1} du (1 - u^2)^{\nu - 1}
\frac{ (1 - z^2) }{[ 1 - 2a(u) z + z^2]^{\nu + 1}}.
\end{equation}
 Performing the external integration first leads to the
simple elementary function, so that the final integral looks like:
\begin{equation}
\label{integral}
J_n = \int_{-1}^{1} du \frac{\sqrt{1 - u^2}}{(u - b)^n}.
\end{equation}
It can be easily evaluated by considering its analytical continuation
to the whole complex plane and choosing the integration contour as
displayed in Fig. \ref{contour}. It reduces to the evaluation of the
residues in the $n$-th order pole in the point $u = b \equiv
(1 - (2x - 1)(2y - 1))/4\sqrt{x \bar x y \bar y}$ and the point
$u = \infty$:
\begin{equation}
J_n =
\pi \left\{ \res_{u = b} + \res_{u = \infty} \right\}
\frac{\sqrt{u^2 - 1}}{(u - b)^n}.
\end{equation}
The first residue provides the contribution proportional to difference
of $\theta$-functions, while the second one --- to the sum. Namely, for
the $n = 1,2$ which appear in the transformation of the one-loop
splitting functions we get
\begin{equation}
\res_{u = b}
\frac{\sqrt{u^2 - 1}}{(u - b)^n}
=
\left\{
\frac{1}{2}\frac{y - x}{\sqrt{x \bar x y \bar y}}
\left[
\theta (y - x) - \theta (x - y)
\right], \qquad\qquad \mbox{for} \quad n = 1
\atop
\frac{1}{2}\frac{1 - (2x - 1)(2y - 1)}{y - x}
\left[
\theta (y - x) - \theta (x - y)
\right], \quad  \mbox{for} \quad n = 2
\right. ,
\end{equation}
and
\begin{equation}
\res_{u = \infty}
\frac{\sqrt{u^2 - 1}}{(u - b)^n}
=
\left\{
-
\frac{1}{4}
\frac{1 - (2x - 1)(2y - 1)}{\sqrt{x \bar x y \bar y}}
\left[
\theta (y - x) + \theta (x - y)
\right], \quad \mbox{for} \quad n = 1
\atop
-
\left[
\theta (y - x) + \theta (x - y)
\right], \qquad\qquad\qquad\quad \mbox{for} \quad n = 2
\right. .
\end{equation}
Thus, we finally obtain
\begin{eqnarray}
J_1 &=& - \frac{\pi}{\sqrt{x \bar x y \bar y}}
\left\{
x \bar y \theta (y - x)
+
y \bar x \theta (x - y)
\right\} , \\
J_2 &=& 2 \pi
\left\{
\frac{x \bar y}{y - x} \theta (y - x)
+
\frac{y \bar x}{x - y} \theta (x - y)
\right\} .
\end{eqnarray}

For instance, for the $QQ$-kernel\footnote{In this section, for brevity,
we omit the colour prefactors in the evolution kernels.} ${^{QQ}\!P}(z)
= (1 + z^2)/(1 - z)$ we have
\begin{equation}
\int_{0}^{1} dz\ {^{QQ}\!P}(z)
\sum_{j = 0}^{\infty}
\frac{w(x | {\scriptstyle\frac{3}{2}})}{N_j(\frac{3}{2})}
C^\frac{3}{2}_j (2x - 1) z^j C^\frac{3}{2}_j (2y - 1)
=
\frac{1}{\pi}
\left\{ \frac{1}{2 y \bar y} J_2
- \sqrt{\frac{x \bar x}{y \bar y}} J_1
\right\} ,
\end{equation}
and after little algebra we come to the known non-singlet kernel.

\subsection{$QG$ and $GQ$ channels.}

We can make use of the same summation formula Eq. (\ref{DGLAP->ER-BL})
for the mixed channels too. For that we have to differentiate both sides
of this identity with respect to the first $x$ or the second $y$
arguments. Since the anomalous dimensions of the conformal operators
differ from the one of the usual local operators by the factor $6/j$ and
$j/6$ for $QG$ and $GQ$ channels, respectively, we will introduce these
factors by representing the forward evolution kernels as a convolution
with appropriate functions. The main reason for that is to use the same
Eq. (\ref{DGLAP->ER-BL}), since otherwise the evaluation of the infinite
series may turn out to be quite complicated. For the $QG$ and $GQ$
sectors we define the kernels
\begin{eqnarray}
{^{QG}\!\widetilde P} (z)
&\equiv& \int_{0}^{1} dx \frac{6}{x}
\int_{0}^{1} dy\ {^{QG}\! P} (z) \delta (z - xy) , \\
{^{GQ}\!\widetilde P} (z)
&\equiv& \int_{0}^{1} dx x^{2\nu - 1}
\int_{0}^{1} dy\ {^{GQ}\! P} (z) \delta (z - xy) ,
\end{eqnarray}
and corresponding summation formulae
\begin{eqnarray}
&&\sum_{j = 1}^{\infty}
\frac{w(x | \nu)}{N_j(\nu)}
C^\nu_j (2x - 1) z^j C^{\nu + 1}_{j-1} (2y - 1)
= 2^{2\nu} \frac{\nu + 1}{\pi} (x \bar x)^{\nu - \frac{1}{2}} \\
&&\qquad\qquad\qquad\times
\int_{-1}^{1} du (1 - u^2)^{\nu - 1}
\left[
(2x - 1) - (2y - 1) \sqrt{\frac{x \bar x}{y \bar y}} u
\right]
\frac{ z (1 - z^2) }{[ 1 - 2a(u) z + z^2]^{\nu + 2}}, \nonumber\\
&&\sum_{j = 1}^{\infty}
\frac{w(x | \nu + 1)}{N_{j - 1}(\nu + 1)}
(2 \nu + j)
C^{\nu + 1}_{j - 1} (2x - 1)
\frac{j}{6} z^j
C^{\nu}_{j} (2y - 1)
= 2^{2\nu} \frac{\nu + 1}{\pi} \frac{(4\nu)^2}{6}
(x \bar x)^{\nu + \frac{1}{2}} \\
&&\qquad\qquad\qquad\times
\int_{-1}^{1} du (1 - u^2)^{\nu - 1}
\left[
(2y - 1) - (2x - 1) \sqrt{\frac{y \bar y}{x \bar x}} u
\right]
\frac{ z (1 - z^2) }{[ 1 - 2a(u) z + z^2]^{\nu + 2}} . \nonumber
\end{eqnarray}
Since the parity odd channel does not introduce any new features as
compared with results available in the literature we turn to the vector
sector. The integrals appeared in the calculation are just the same as
before, namely, we get for the ${^{QG}\!P}^V (z) = (z^2 + (1-z)^2)$
splitting function
\begin{eqnarray}
&&\int_{0}^{1} dz\ {^{QG}\!\widetilde P}^V (z)
\sum_{j = 1}^{\infty}
\frac{w(x | {\scriptstyle \frac{3}{2}})}{N_j(\frac{3}{2})}
C^\frac{3}{2}_j (2x - 1) z^j C^\frac{5}{2}_{j-1} (2y - 1) \\
&&\hspace{4cm} = \frac{1}{\pi}
\left\{
\frac{x - y}{4(y \bar y)^2} J_2
- \sqrt{\frac{x \bar x}{y \bar y}}
\frac{4x - 2y - 1}{2 y \bar y} J_1
+ 4 \frac{x \bar x}{y \bar y} (2 y - 1) J_0
\right\} , \nonumber
\end{eqnarray}
and for the ${^{GQ}\!P}^V (z) = (1 + (1 - z)^2)/z$ kernel
\begin{eqnarray}
&&\int_{0}^{1} dz\ {^{GQ}\!\widetilde P}^V (z)
\sum_{j = 1}^{\infty}
\frac{w(x | {\scriptstyle \frac{5}{2}})}{N_{j - 1}(\frac{5}{2})}
(3 + j)
C^\frac{5}{2}_{j - 1} (2x - 1)
\frac{j}{6} z^j
C^\frac{3}{2}_j (2y - 1) \\
&&\hspace{4cm} = \frac{1}{\pi}
\left\{
\frac{y - x}{4 y \bar y} J_2
- \frac{1}{2} \sqrt{\frac{x \bar x}{y \bar y}} (4y - 2x - 1) J_1
+ 4 x \bar x (2x - 1) J_0
\right\} . \nonumber
\end{eqnarray}
Discarding the constant term $J_0 = \frac{\pi}{2}$ we reproduce the
Chase's kernels, however, if the latter is taken into account we come
to the expressions given in Eqs. (\ref{improved}).

\section{Summary.}

In the present paper we have developed a formalism, which allows to
evaluate the $n$-th order non-diagonal part of the exclusive twist-2
evolution kernels starting form the $( n - 1 )$-loop approximation
for the special conformal anomaly matrix, for the flavour singlet
channel in Abelian gauge theory. The main tools of our analysis are
the renormalized anomalous conformal Ward identities and commutator
constraint obtained from the algebra of the conformal group. Using
this approach we have derived ${\cal O} (\alpha^2)$ corrections to
singlet anomalous dimensions of the conformal operators. This is
sufficient to reconstruct the complete ER-BL splitting functions
${^{AB}V^{(1)}} ( x, y )$ in the NLO approximation. We have partially
checked our calculations by explicit evaluation of the simple two-loop
diagrams with fermion vacuum polarization insertions. The formalism we
have described allows to write in a simple way the solution of the
corresponding evolution equations, since the corrections to the
conformal partial wave expansion are defined by known one-loop
evolution kernels and $g$-functions \cite{Mul94,BM97a}. We have also
derived, for the first time, the reduction formulae which allow to
find the diagonal part of the ER-BL kernels from the corresponding
DGLAP analogues. This procedure fix unambiguously the form of the
diagonal part of the exclusive kernels which are exactly diagonal
in the physical as well as in non-physical spaces of moments.

\vspace{0.5cm}

We wish to thank B. Geyer, A.V. Radyushkin, D. Robaschik for valuable
correspondence. A.B. was supported by the Alexander von Humboldt
Foundation and partially by Russian Foundation for Fundamental
Research, grant N 96-02-17631.

\vspace{0.5cm}

\appendix

\setcounter{section}{0}
\setcounter{equation}{0}
\renewcommand{\theequation}{\Alph{section}.\arabic{equation}}

\section{Feynman rules.}

In this appendix we summarize Feynman rules which have been used in
the calculation of the renormalization constants in the main text.
Everywhere it is implied that all momenta on the lines are incoming.

For quark and gluon non-local string operators we have
\begin{eqnarray}
\left\{\!\!\!
\begin{array}{c}
{^G\!{\cal O}^V} \\
{^G\!{\cal O}^A}
\end{array}
\!\!\!\right\}_{\mu\nu} \!\!\!\!&=&\!\!\!\!
\left\{\!\!\!
\begin{array}{c}
g_{\alpha\beta} \\
i \epsilon_{\alpha\beta -+}
\end{array}
\!\!\!\right\}
f_{+ \alpha; \nu} (k_2) f_{+ \beta; \mu} (k_1)
\left[
e^{ - i \kappa_1 k_{1+} - i \kappa_2 k_{2+} }
\pm
e^{ - i \kappa_1 k_{2+} - i \kappa_2 k_{1+} }
\right] , \\
\left\{\!\!\!
\begin{array}{c}
{^Q\!{\cal O}^V} \\
{^Q\!{\cal O}^A}
\end{array}
\!\!\!\right\}
&=&\!\!\!\!
\left\{\!\!\!
\begin{array}{c}
\gamma_+ \\
\gamma_+ \gamma_5
\end{array}
\!\!\!\right\}
\left[
e^{ - i \kappa_1 k_{1+} - i \kappa_2 k_{2+} }
\mp
e^{ - i \kappa_1 k_{2+} - i \kappa_2 k_{1+} }
\right] .
\end{eqnarray}
In the quark sector we have omitted the contributions coming
form the phase factor, since they are not of relevance for our
present consideration.

The $\W$-operator in parity-even and -odd channels read,
respectively,
\begin{eqnarray}
\left\{\!\!\!
\begin{array}{c}
\W^V
\\
\W^A
\end{array}
\!\!\!\right\}_{\mu\nu}
\!\!\!\!&=&\!\!\!\! 2 i
\left\{\!\!\!
\begin{array}{c}
g_{\alpha\beta}
\\
i \epsilon_{\alpha\beta - +}
\end{array}
\!\!\!\right\}
\left\{
( g_{\beta \mu} - n_{\beta}^* n_{\mu} ) f_{+ \alpha; \nu} (k_2)
+
( g_{\alpha \nu} - n_{\alpha}^* n_{\nu} ) f_{+ \beta; \mu} (k_1)
\right\} \nonumber\\
&&\hspace{5cm}\times
\left[
e^{ - i \kappa_1 k_{1+} - i \kappa_2 k_{2+} }
\pm
e^{ - i \kappa_1 k_{2+} - i \kappa_2 k_{1+} }
\right] ,
\end{eqnarray}
where $f_{\alpha \beta; \mu} (k) \varepsilon_\mu = k_\alpha
\varepsilon_\beta - k_\beta \varepsilon_\alpha$ is the strength
tensor in the momentum space.

The only modified Feynman rule coming from the differential
operator insertion $[\Delta^\g_-]$, required to the order we
are interested in, is
\begin{equation}
\label{modifiedFR}
\widetilde\V_\mu = i \g \gamma_\mu (2 \pi)^4
2 i \partial_- \delta^{(4)} (k_1 + k_2 + k_3).
\end{equation}
It should be compared to the usual one for the quark-gluon vertex
\begin{equation}
\label{usualFR}
\V_\mu = i \g \gamma_\mu (2 \pi)^4
\delta^{(4)} (k_1 + k_2 + k_3).
\end{equation}
The resulting expression, composed from propagators and vertices
we have just derived, should be integrated with respect to the
momentum of every internal line, i.e. multiplied by the factor
\begin{equation}
\int \prod_\ell \frac{d^dk_\ell}{(2\pi)^{\ell\cdot d}}.
\end{equation}

\setcounter{equation}{0}
\section{Generalized ER-BL kernels.}

Recently, there is renewed interest in the deeply virtual
Compton scattering \cite{MRGDH94,Ji97-DVCS,Radyushkin97,Ji97}. The
evolution of the corresponding non-forward parton distribution
functions \cite{Radyushkin97}
\begin{eqnarray}
\label{G-definition}
\langle h' |
{^G\!{\cal O}^{\mit\Gamma}} (\kappa_1 , \kappa_2)
|h \rangle
&=& \frac{1}{2} \int d x
\left[
e^{ - i \kappa_1 x - i \kappa_2 ( \zeta - x ) }
\pm
e^{ - i \kappa_2 x - i \kappa_1 ( \zeta - x ) }
\right]
{^G\!{\cal O}^{\mit\Gamma}} (x, \zeta), \\
\langle h' |
{^Q\!{\cal O}^{\mit\Gamma}} (\kappa_1 , \kappa_2)
|h \rangle
&=& \int d x
\left[
e^{ - i \kappa_1 x - i \kappa_2 ( \zeta - x ) }
\mp
e^{ - i \kappa_2 x - i \kappa_1 ( \zeta - x ) }
\right]
{^Q\!{\cal O}^{\mit\Gamma}} (x, \zeta) ,
\end{eqnarray}
where the non-local string operators are defined by Eqs.
(\ref{string-operator}) and the upper (lower) signs corresponds to the
parity-even (odd) channels, is governed by generalized ER-BL evolution
equations
\begin{eqnarray}
\label{MOMEE}
\mu^2\frac{d}{d\mu^2}{\cal O}^{\mit\Gamma} (x, \zeta) &=&
- \frac{\alpha}{2 \pi} \int d x ' K^{\mit\Gamma} (x , x ', \zeta)
{\cal O}^{\mit\Gamma} (x ', \zeta).
\end{eqnarray}
We perform the calculations and end up with ER-BL type kernels which
look like for the parity-even sector
\begin{eqnarray}
\label{PE-ext}
&&{^{QQ}\!K^V} (x , x ', \zeta) \\
&&\quad\qquad = C_F \,
\left[ \frac{x}{x - x'}
\Theta_{11}^0 (x , x - x ') + \frac{x - \zeta}{x - x'}
\Theta_{11}^0 (x - \zeta, x - x ')
+ \Theta_{111}^0 ( x , x - \zeta, x - x ' )
\right]_+ , \nonumber\\
&&{^{QG}\!K^V} (x , x ', \zeta) = 2 T_F N_f \,
\left[ \Theta_{112}^1 (x, x - \zeta , x - x ')
- 2\frac{x - x'}{x' (x' - \zeta)}
\Theta_{111}^0 (x, x - \zeta, x - x ')
\right] , \nonumber\\
&&{^{GQ}\!K^V} (x , x ', \zeta) = C_F \,
\left[ ( 2 x' - \zeta ) \Theta_{111}^0 (x, x - \zeta , x - x ')
- x (x - \zeta)
\Theta_{111}^1 (x, x - \zeta, x - x ')
\right] , \nonumber
\end{eqnarray}
and for parity odd one
\begin{eqnarray}
&&{^{QQ}\!K^A} (x , x ', \zeta)
= {^{QQ}\!K^V} (x , x ', \zeta), \\
&&{^{QG}\!K^A} (x , x ', \zeta) = 2 T_F N_f \,
\Theta_{112}^1 (x, x - \zeta , x - x ') , \nonumber\\
&&{^{GQ}\!K^A} (x , x ', \zeta) = C_F \,
\left[ ( 2 x - \zeta ) \Theta_{111}^0 (x, x - \zeta , x - x ')
- x (x - \zeta)
\Theta_{111}^1 (x, x - \zeta, x - x ')
\right] . \nonumber
\end{eqnarray}
We have used here the generalized $\Theta$-functions defined by
the integral
\begin{equation}
\Theta^{m}_{i_1 i_2 ... i_n}
(x_1,x_2,...,x_n)=\int_{-\infty}^{\infty}\frac{d\alpha}{2\pi i}
\alpha^m \prod_{k=1}^{n}\left(\alpha x_k -1 +i0 \right)^{-i_k}.
\end{equation}
Using the identities
\begin{eqnarray}
\Theta^n_{ i j k \cdots } (x_1, x_2, x_3, \dots)
&=& \frac{1}{x_1 - x_2}
\left\{
\Theta^{n - 1}_{ i - 1 j k \cdots } (x_1, x_2, x_3, \dots)
-
\Theta^{n - 1}_{ i j - 1 k \cdots } (x_1, x_2, x_3, \dots)
\right\} \\
&=& \frac{1}{x_1 - x_2}
\left\{
x_2 \Theta^n_{ i - 1 j k \cdots} (x_1, x_2, x_3, \dots)
-
x_1 \Theta^n_{ i j - 1 k \cdots} (x_1, x_2, x_3, \dots)
\right\}, \nonumber
\end{eqnarray}
we can always reduce them to the lowest ones which have the following
explicit form
\begin{eqnarray}
&&\Theta^0_1 (x) = 0,\\
&&\Theta^0_2 (x) = \delta (x),\\
&&\Theta^0_{11} (x_1,x_2)
=\frac{\theta(x_1)\theta(-x_2)
-\theta(x_2)\theta(-x_1)}{x_1-x_2} .
\end{eqnarray}

Again the parity even kernels (\ref{PE-ext}) possess the off-diagonal
pieces in the non-physical sector. To cure this problem in appropriate
way, so that the forward limit would not be affected, we should add the
terms
\begin{equation}
\Delta K^V (x , x ', \zeta) =
2 \Theta_{11}^0 (x , x - \zeta) \frac{x (\zeta - x)}{x' (\zeta - x')}
\times
\left\{
{( \zeta - 2 x' ),\qquad\qquad\quad \mbox{for $QG$-channel}}
\atop
{x' (\zeta - x')( \zeta - 2x ), \quad \mbox{for $GQ$-channel}}
\right. .
\end{equation}
In the forward limit $\Theta_2^0 (x) = \delta (x)$, so that these extra
terms die out.

As usual these kernels possess the following DGLAP and ER-BL limits:
\begin{equation}
P (z) \equiv - K (z , 1, 0), \qquad
V ( x, y ) \equiv - K (x , y, 1) .
\end{equation}

\end{document}